\documentclass[10pt]{article}

\usepackage[superscript,sort]{cite}

\usepackage{ifthen,ifpdf}
\ifpdf
	\usepackage[pdftex]{hyperref}	 \hypersetup{colorlinks=true,linkcolor=black,citecolor=black,urlcolor=blue,backref=page,bookmarks=true,breaklinks=true,plainpages=false }
	\usepackage[pdftex]{graphicx}
	\usepackage[usenames,pdftex]{color}
\else
	\usepackage[colorlinks=true,breaklinks=true,plainpages=false]{hyperref}
	\usepackage{graphicx}
	\usepackage[usenames]{color}
\fi

\usepackage{amsthm,amscd,amsxtra,amsfonts,amsmath,amssymb,multirow}
\usepackage{wrapfig}
\usepackage[footnotesize]{caption}
\usepackage[tiny,compact]{titlesec}
\usepackage[textwidth=0.8in,textsize=footnotesize]{todonotes}

\usepackage{longtable}
\usepackage{multirow}
\usepackage{algorithm}
\usepackage[noend]{algpseudocode}

\usepackage{helvet}

\usepackage[letterpaper,top=0.5in,bottom=0.5in,left=0.5in,right=0.5in]{geometry}

\titlespacing*{\section}{0pt}{*0}{*0}
\titlespacing*{\subsection}{0pt}{*0}{*0}
\titlespacing*{\subsubsection}{0pt}{*0}{*0}
\titlespacing{\paragraph}{0pt}{*0}{*1}
\definecolor{MyPurple}{rgb}{1,0,1}

\begin{document}

%\pagenumbering{roman}
%\begin{verbatim}

\title{Feature functional theory - binding predictor (FFT-BP) for the blind prediction of binding free energies}

\author{Bao Wang$^{1}$\footnote{Current address: Box 951555
Los Angeles, CA 90095-1555.}, Zhixiong Zhao$^{2}$\footnote{The first two authors contribute equally to this work.}, Duc D. Nguyen$^{1}$ and Guo-Wei Wei$^{1,3,4}$
\footnote{
Correspondences should be addressed to Guo-Wei Wei. E-mail: wei@math.msu.edu.}  \\
$^1$Department of Mathematics \\
Michigan State University, MI 48824, USA\\
$^2$School of Medicine, Foshan University, Foshan, Guangdong 528000, China\\
$^3$Department of Electrical and Computer Engineering \\
Michigan State University, MI 48824, USA \\
$^4$Department of Biochemistry and Molecular Biology\\
Michigan State University, MI 48824, USA
}

\date{\today}

\maketitle

\begin{abstract}
We present a   feature functional theory - binding predictor (FFT-BP) for the  protein-ligand binding affinity prediction.  The  underpinning assumptions of FFT-BP  are as  follows: i) representability: there exists a  microscopic feature vector that can uniquely characterize   and distinguish one  protein-ligand complex  from another; ii) feature-function relationship: the macroscopic features, including binding free energy,  of a complex is a functional of microscopic feature vectors; and iii) similarity:  molecules with similar  microscopic  features have similar  macroscopic features, such as binding affinity. Physical models, such as implicit solvent models and quantum theory, are utilized to extract microscopic features, while  machine learning algorithms are employed to rank the similarity among protein-ligand complexes. A large variety of numerical validations and  tests confirms the accuracy and robustness of the proposed FFT-BP model. The root mean square errors (RMSEs) of FFT-BP blind predictions of  a benchmark set of 100 complexes,  the PDBBind v2007 core set of 195 complexes and the PDBBind v2015 core set of 195 complexes are 1.99, 2.02 and 1.92 kcal/mol, respectively. Their corresponding Pearson correlation coefficients are  0.75, 0.80,  and 0.78, respectively.

\end{abstract}

\newpage

 {\setcounter{tocdepth}{4} \tableofcontents}
\newpage

 { 
\begin{center}
	\begin{longtable}{ll}	
		\caption{List of abbreviations and symbols}\label{tab:symbols}\\
			\hline \multicolumn{1}{l}{Sign} &\multicolumn{1}{l}{Description} \\ \hline
			\endfirsthead
			{{\bfseries \tablename\ \thetable{} -- \footnotesize{continued}}} \\
			\hline \multicolumn{1}{l}{Sign} &\multicolumn{1}{l}{Description} \\ \hline
			\endhead
			\endfoot
				\hline
			\endlastfoot
			BRT 	& Boosted regression trees\\
			CAAD 	& Computer aided drug design \\
			EM 	& expectation-maximization algorithm\\
			ESES 	& Eulerian solvent excluded surface software\\ 
			FFT 	& Feature functional theory \footnote{Please do not confuse with fast Fourier transform.}\\
			FFT-BP 	& Feature functional theory - binding predictor\\
			GA 	& Genetic algorithm\\
			GBDT 	& Gradient boosting decision tree\\
			KECSA 	& Knowledge-based and empirical combined scoring algorithm\\
			kNN 	& k-nearest neighbours\\
			LJ 	& Lennard Jones\\
			LTR 	& Learn to rank\\
			MARS 	& Multivariate adaptive regression\\
			MART 	& Multiple additive regression tree\\
			MC 	& Monte Carlo\\
			MD 	& Molecular dynamics \\
			MIBPB 	&  Matched interface and boundary-based Poisson-Boltzmann equation solver\\
			MLR 	& Multiple linear regression \\
			MM GBSA 	& Molecular mechanics generalized-Born surface area\\
			MM PBSA 	& Molecular mechanics Poisson-Boltzmann surface area\\
			PB 	& Poisson-Boltzmann\\
			PMF 	& Potential of the mean force\\
			QM/MM 	& Quantum mechanics/molecular mechanics \\
			QSAR 	& Quantitative structure-activity relation\\
			RF 	& Random forest\\
			RMSE 	& Root mean square error\\
			SPT 	& Scaled-particle theory\\
			SVR 	& Support vector regression \\
			vdW 	& van der Waals\\
			VS 	& Virtual screening \\
			$[*]$& Jump of quantity $*$ across the interface $\Gamma$\\
			$\|\mathbf{r}_i - \mathbf{r}_j\|$ & Distance between two points located at $\mathbf{r}_i$ and $\mathbf{r}_j$\\
			$\Gamma$ & Solvent-solute interface \\
			$\Delta E_{\text{MM}}$ & Molecular mechanics energy\\
			$\Delta G$ & Binding free energy\\
			$\Delta G_{\rm AB}$ & Binding free energy of molecular complex AB\\
			$\Delta G_i$ & Binding free energy for $i$th complex\\
			$\Delta G_{\text{el}}$ & Electrostatics binding free energy between protein and ligand\\
			$\Delta G_{\text{Coul}}$ & Coulombic interaction\\
			$\Delta G_{\text{RF}}$ & Reaction field energy\\
			$\Delta G_{\text{RF}_i}$ & Atomic reaction field for $i$th atom\\
			$(\Delta G_{\text{RF}})_\text{Com}$ & Reaction field energy of complex\\
			$(\Delta G_{\text{RF}})_\text{Lig}$ & Reaction field energy of ligand\\
			$(\Delta G_{\text{RF}})_\text{Pro}$ & Reaction field energy of protein\\
			$\Delta G_{\text{solv}}$ & Solvation free energy\\
			$\delta(\mathbf{r}-\mathbf{r}_i)$ & Delta function at point $\mathbf{r}_i$\\
			$\phi$ & Electrostatics potential\\
			$\phi_\text{h}$ & Electrostatics potential in homogeneous environment\\
			$\phi_{\mathbf{n}}$ & Normal derivative of function $\phi(\mathbf{r})$\\
			$\epsilon(\mathbf{r})$ & permittivity function\\
			$\epsilon_{ij}$ & Measurement of the depth of the attractive well in van der Waals interaction $u_{ij}$ \\
			$\epsilon_m$ & Dielectric value of solute domain\\
			$\epsilon_s$ & Dielectric value of  solvent domain\\
			$\Omega$ & $\Omega^{\rm m}\bigcup \Omega^{\rm s}$ \\
			$\Omega^{\text{m}}$ & Solute domain\\
			$\Omega^{\text{s}}$ & Solvent domain\\
			$f_{\rm binding}$ & Unknown functional for modelling the relationship between binding free energy and extended features\\
			$f_j$ & unknown function modeling the $j$th physical observable of molecule A\\
			$N_m$ & Number of atoms in molecule \\
			${\bf o}_{{\rm A }j}$ & $j$th physical observable ${\bf o}_j$ of target molecule A\\
			$\mathbf{o}_{ij}$ & $j$ macroscopic feature for $i$th molecule or complex\\
			$\mathbf{o}_i$ & Macroscopic feature vector for $i$th molecule or complex\\
			$Q_i$ & Partial charge located at $\mathbf{r}_i$\\
			$\mathbf{r}$ & Vector in $\mathbf{R}^3$\\
			$\mathbf{r}_i$ & 3D coordinate of $i$th atom\\
			$r_i$ & Atomic radius of the $i$th atom\\
			$r_{ij}$ & Distance between two points located at $\mathbf{r}_i$ and $\mathbf{r}_j$\\
			$T\Delta S$ & Entropy\\
			$u_{ij}$ & van der Waals interaction between $i$th and $j$th atoms\\
			$\mathbf{v}_i$ & Extended feature vector $\mathbf{v}_i=(\mathbf{x}_i,\mathbf{o}_i)$ for $i$th molecule or complex\\
			$\mathbf{x}_{\rm A}$ & Microscopic feature vector of the target molecule A\\
			$\mathbf{x}_{\rm AB}$ & Microscopic feature vector of the target complex AB\\
			$\mathbf{x}_i$ & Microscopic feature vector for $i$th molecule or complex\\
			$\mathbf{x}_{ij}$ & $j$th microscopic feature for $i$th molecule\\
%			\hline
	\end{longtable}
\end{center}
}

\section{Introduction}
Designing efficient drugs for curing diseases is of essential importance for the new century's life science. Indeed, one of the ultimate goals of molecular biology is to understand the molecular mechanism of human diseases  and  to develop efficient side-effect-free drugs for  disease curing. Nevertheless, the drug discovery procedure is extremely complicated, and involves many scientific disciplines and technologies.  As a brief summary, the drug discovering contains the following seven major steps \cite{Bock:2002}, namely, i)  Disease identification; ii) Target hypothesis, i.e., the activation or inhibition of drug targets (usually proteins within the cell)  is thought to alter the disease state; iii) Screening potential principle compounds that will bind to the target; iv) Optimizing the identified compounds with respect to their structural characteristics in the context of the target binding site; v)  Preclinical test, both \textit{in vitro} and \textit{in vivo} tests will be performed; vi) Clinical trials to determine their bioavailability and therapeutic potential; and vii) Optimizing chemical's efficacy, toxicity, and pharmacokinetics properties. Typically, the whole cost of a new drug development is estimated to be more than one billion dollars with more than ten years' research efforts \cite{WeiZhang:2015}. Such large amount of cost is mostly due to  unsuitable chemical compounds that are used in the preclinical and clinical testing \cite{Shivani:2010}. In terms of economical drug design, sophisticated and accurate computer aided compound screening methods become extremely important. Virtual screening (VS) methodologies focus on detecting a small set of highly promising candidates for further experimental testing \cite{VirtualScreeningShoichet:2004}. Docking is one of the most important VS methodologies and is widely used in the computer aided drug design (CADD). It is a two-stage protocol \cite{Ballester:2014}. The first step is the sampling of the ligand binding conformations, which determines the pose, orientation, and conformation of a molecule as docked to the target binding site \cite{Bursulaya:2003}. The second stage is the scoring of protein-ligand binding affinity. With the  development of molecular dynamics (MD), Monte Carlo (MC), and genetic algorithm (GA) for pose generation, the sampling problem is relatively well resolved \cite{RenxiaoWang:2003CompareSF, Leach:2006, Novikov:2011}. A major remaining challenge  in achieving accurate docking is the development of accurate scoring functions for diverse protein ligand complexes. One of the most important open problems in computational biosciences is the accurate prediction of the binding affinities of a large set of diverse protein-ligand complexes \cite{Ballester:2014}. A desirable goal is to achieve less than 1 kcal/mol   root mean square error (RMSE) in the prediction.

Since the pioneer work in the   1980s and 1990s, the study of the scoring function and sampling techniques has been blooming in the CADD community \cite{Kuntz:1982,DesJarlais:1986,Goodsell:1990,Jorgensen:1991}. In a recent review, Liu and Wang classify the existing popular scoring functions into four categories   \cite{LiuJie:2014}, namely, i) Force-field based or physical based scoring functions; ii) Empirical or regression based scoring functions; iii) Potential of the mean force (PMF) or knowledge based scoring functions; and iv) Machine learning based scoring functions. Physics based scoring functions provide some of the most accurate and detailed description of the protein and ligand molecules in the solvent environment. Typical models that belong to this category are molecular mechanics Poisson-Boltzmann surface area (MM PBSA) and molecular mechanics generalized-Born surface area (MM GBSA) \cite{Kollman:2000,Gilson:2007} with a given force field parametrization of both solvent and solute molecules, like Amber or CHARMM force fields \cite{Wang:2004b, CHARMM22, Weiner:1986}. In this framework, the binding free energy is often modeled as a superposition of four parts: van der Waals (vdW), electrostatics interactions between protein and ligand, the hydrogen bonding, and solvation effects.  In addition to MM PBSA and MM GBSA, several other prestigious scoring functions also belong to this category, including COMBINE \cite{Ortiz:1995} and MedusaScore \cite{Yin:2008}. Physical based scoring functions are a class of dynamically improved methods and the VS can become more and more accurate with the further development of  advanced and comprehensive molecular mechanics force fields. Plenty of improvement has already been done for improving the accuracy of scoring functions, such as  QM/MM multiscale coupling \cite{Su:2015} and polarizable force fields \cite{Ponder:2010}. Empirical or regression based scoring functions, usually also called multiple linear regression (MLR) scoring functions, typically model the protein-ligand binding affinity contributed from vdW interaction, hydrogen bonding, desolvation, and metal chelation \cite{Zheng:2015LISA}. Several parameters are introduced in each of the above term, and the scoring function is obtained by using the existing protein-ligand binding information to train these parameters in the given binding affinity function. Many other existing scoring functions also belong to this category, e.g., PLP \cite{Verkhivker:1995PLP}, ChemScore \cite{Eldridge:1997}, and X-Score \cite{WangRenXiao:2002}, etc.

A recent study on a congeneric series of thrombin inhibitors concludes that free energy contributions to protein-ligand binding are non-additive, showing some  theoretical  deficiencies of the MLR based scoring functions \cite{Baum:2010}. The theoretical basis of this non-additivity was explained in an earlier review \cite{zhou2009theory}. Machine learning algorithms do not explicitly require a given form of the binding affinity to its related items, and thus do not require the additive assumption of energetic terms. Many machine learning based scoring functions are proposed in the past few decades. These methods apply quantitative structure-activity relation (QSAR) principles to the prediction of the protein-ligand binding affinity. Representative work along this line is the random forest (RF) based scoring function, RF-Score \cite{HongjianLi:2014RF}. In RF-Score, the random forest is selected as the basic regressor instead of the classical MLR, which is restricted to the pre-defined linear form of the binding affinity function. By utilization of the features calculated from the existing scoring functions, it achieves highly accurate five-fold cross validation results on the PDBBind v2013 refined set. Prediction results on the PDBBind  v2007  core set further confirms the accuracy of the RF-Score \cite{HongjianLi:2014RF}. Many other machine learning tools are utilized as the main  skeletons  of the scoring functions, like support vector regression (SVR) \cite{Sarah:2011}, multivariate adaptive regression (MARS), k-nearest neighbours (kNN), boosted regression trees (BRT), etc \cite{Ashtawy:2012}. The blooming of the big data approaches and more accurate descriptors characterization of the protein-ligand binding effects have made machine learning type of scoring functions full of vitality in CADD. Machine learning based scoring functions can make continuous improvement through both  advance  in physical protein-ligand binding descriptors and  discovery of new machine learning techniques.

Another important class of scoring functions is PMF based. This category of scoring functions is based on the simplified statistical mechanics theory in which the protein ligand binding affinity is modeled as the sum of pairwise statistical potentials between protein and ligand atoms. The major merit of the PMF type of scoring functions is their simplicity in both concept and computation. This simplified physical model captures major physical principles behind the protein ligand binding. In knowledge-based and empirical combined scoring algorithm (KECSA), the binding affinity between protein and ligand are modeled by 49 pairwise  modified  Lennard-Jones types of potentials between different types of atoms \cite{Zheng:2013KECSA}. Through a large number of training instances, the functional form of all these pairwise interaction potentials can be determined.  Effective ligand binding conformation   sampling procedure can also be incorporated into this theoretical framework \cite{Zheng:2013MoveableType}. There are also many other interesting developments in the PMF based scoring functions, e.g.,  PMF \cite{PMFScore:1999}, DrugScore \cite{DrugScore:2005}, and IT-Score \cite{ITScore:2006}.

Essentially, the major purpose of a scoring function is to find the relative order of binding affinities of   candidate chemicals  to the target binding site. This ranking result is  further used  for the preclinical test in a realistic drug design procedure. From this point of view, the development of scoring functions turns out to be the development of ranking methods. Many existing scoring functions have been developed from this perspective.  For example,  learning to rank (LTR) algorithms have been used to develop various scoring functions, including PTRank, RankNet, RankNet, RankBoost, ListNet, and AdaRank \cite{WeiZhang:2015, Wassermann:2009, Shivani:2010, Wale:2009}. Compared to  other machine learning or simple MLR based scoring functions, the advantages of  ranking based scoring functions are two-fold.  First, they are applicable to identifying compounds on novel protein binding sites where no sufficient data available for  other machine learning algorithms. Second, they are suitable for the case that binding affinities are measured in different platforms since ranking can be more focused on  relative order \cite{WeiZhang:2015}.

In this work, we propose a feature functional theory-binding predictor (FFT-BP) for the blind prediction of binding affinity. The FFT-BP is constructed based on three assumptions, i.e., i) representability assumption: there exists a  microscopic feature vector that can uniquely characterize   and distinguish one  molecular complex from another; ii) feature-function relationship assumption: the macroscopic features, including binding free energy,  of a molecule or complex is a functional of microscopic feature vectors; and iii) similarity assumption:  molecules or complexes with similar  microscopic  features have similar  macroscopic features, such as binding free energies. FFT-BP has three distinguishing traits. A major trait of the proposed FFT-BP is its use of microscopic features derived from physical models, including Poisson Boltzmann (PB) theory \cite{Sharp:1990, Sharp:1990a, Rocchia:2001, Honig:1995a, Gilson:1993, ZhanChen:2010a,BaoWang:2015a}, nonpolar solvation models \cite{Stillinger:1973,Gallicchio:2002,Gallicchio:2004,choudhury2005mechanism,Wagoner:2006,Wei:2009,ZhanChen:2012},  components in MM PBSA  \cite{Kollman:2000} and quantum models.  As such, electrostatic solvation free energy, electrostatic binding affinity,  atomic reaction field energies, and  Coulombic interactions are utilized to represent the electrostatic effects of protein-ligand binding.   Atomic pairwise van der Waals interactions are employed to model the dispersion interactions between the protein and ligand.  We also make use of atomic surface areas and  molecular volume in our FFT-BP to describe hydrophobic and entropy effects of the protein-ligand binding process. Another trait of the present FFT-BP is its feature-function relationship assumption, which avoids the use of additive modelling of the total binding affinity based on the direct sum of various energy components.  Machine learning algorithms  automatically rank the relative importance of various features to the binding affinity. By  utilizing the boosted regression tree type of algorithms for the ranking, our model can capture the nonlinear dependence of the binding affinity to each feature. The other trait  of FFT-BP is its use of advanced LTR algorithm, the multiple additive regression tree (MART), for ranking the nearest neighbors via microscopic features. This approach allows us to further improve our method by incorporating the state-of-the-art  machine learning techniques.

%Structure of this paper
This paper is structured as  follows. In Section \ref{Theory}, we present the theoretical background of FFT-BP,  which consists of four parts, basic assumptions, microscopic feature selection, MART algorithm and binding affinity function. In Section \ref{Results}, we verify the accuracy and robustness of our FFT-BP by a  validation set, a training set  and three standard test sets involving  a  variety of diverse protein-ligand complexes.  We show that FFT-BP delivers some of the best binding affinity predictions.   This paper ends with concluding remarks.

\section{Theory and algorithm} \label{Theory}

In this section, we present FFT for binding free energy prediction. First, we discuss the basic FFT assumptions. Additionally, feature selections are based on physical models. Moreover, protein-ligand complexes are ranked from a machine learning algorithms, i.e., the MART ranking algorithm.  Finally, we describe a prediction algorithm for  approximating  the binding free energy   based on features from nearest neighbors ranked by the MART algorithm.

\subsection{Basic assumptions}\label{Assumption}
Our FFT is based on three assumptions, including representability, feature-functional relationship and similarity. These assumptions are described below.

\paragraph{Representability assumption}
Without lost of generality,  we consider a total of $N$ molecules  or  complexes $\{ M_i\}_{i=1}^N$ with known  names and geometric structures from related databases.
One of FFT basic assumptions is that  there exists an $n$-dimensional microscopic feature vector, denoted as ${\bf x}_i=({\bf x}_{i1}, {\bf x}_{i1},\cdots, {\bf x}_{in})$ to uniquely characterize and distinguish the $i$th molecule or complex. Here the vector components include various microscopic features, such as atomic types and numbers, atomic charges,  atomic dipoles, atomic quadrupole, atomic reaction field energies, electrostatic solvation or electrostatic binding free energies, atomic surface areas, pairwise atomic van der Waals interactions, etc.

For $i$th molecule or complex, apart from its $n$ microscopic features, there are $l$ macroscopic features, or physical observable ${\bf o}_i=({\bf o}_{i1}, {\bf o}_{i1},\cdots, {\bf o}_{il})$, such as density,   boiling point, enthalpy of formation,  heat of combustion,  solvation free energy, pKa,   viscosity, permittivity, electrical conductivity, binding free energy, etc. We combine the microscopic and macroscopic feature vectors to construct  an extended feature vector ${\bf v}_{i}=({\bf x}_i, {\bf o}_i)$ for the $i$th molecule.

Extended feature vectors $\{{\bf v}_i\}_{i=1}^N $ span a  vector space ${\cal V}$, which satisfies commonly required eight axioms for addition and multiplication, such as  associativity, commutativity, identity element,  and   inverse elements of addition,  compatibility of scalar multiplication with field multiplication, etc.
%Identity element of scalar multiplication , Distributivity of scalar multiplication with respect to vector addition, Distributivity of scalar multiplication with respect to field addition.
Unlike the usual $L_p$ space,  the extended feature space does not have the notion of nearness, angles or distances. We therefore need additional  techniques, namely,  machine learning algorithms, to study the nearness and distance between feature vectors.  The selection of microscopic features depends on what physical or chemical prediction is interested. In our approach, we utilize microscopic features form related physical models. For example, for solvation and binding free energy prediction, we select features that are derived from implicit solvent models and quantum mechanics.

Based on our assumption,  microscopic features along are able to characterize and distinguish molecules. In contrast, macroscopic features are
used  as the label in learning and ranking  for a given purpose. Therefore, for a given task, say binding free energy prediction,
we do not include all the macroscopic features in the feature vector ${\bf o}_i$.  We  only select  ${\bf o}_i=({\bf o}_{i1})=\Delta  G_i, \forall i=1,\cdots,N$,
where  $\{\Delta   G_i\}$ are known binding free energies from databases.  The resulting extended vector is used for the binding free energy prediction.
%For this reason, the selection of macroscopic features is described in the dataset preparation, i.e., Section \ref{Sec:Dataset}.

\paragraph{Feature-function relationship  assumption}
In FFT,  a general feature-function relationship is assumed  for the $j$th physical observable ${\bf o}_j$ of target molecule A
\begin{equation}
\label{SolvationEnergyFunctionAssumption2}
{\bf o}_{{\rm A }j}=f_{j}(\mathbf{x}_{\rm A},\mathbf{v}_{1}, \mathbf{v}_{2}, \cdots, \mathbf{v}_{N}),
\end{equation}
where $f_{j}$ is an unknown function modeling the $j$th physical observable of molecule A and  $\mathbf{x}_{\rm A}$ is the microscopic feature vector of the target molecule A. This relation applies to the prediction of various physical and chemical properties.
In the present application, we are interested in the prediction of binding free energies for a set of diverse protein-ligand complexes. We construct a feature space for the training set and the binding free energy of target molecular complex AB can be given as a functional of extended feature vectors
\begin{equation} \label{SolvationEnergyFunctionAssumption}
\Delta  { G}_{\rm AB}=f_{\rm binding}(\mathbf{x}_{\rm AB}, \mathbf{v}_{1}, \mathbf{v}_{2}, \cdots, \mathbf{v}_{N})
\end{equation}
where $\Delta  { G}_{\rm AB}$ is the binding free energy of  molecular complex AB, and $f_{\rm binding}$ is an unknown functional for modeling the
relationship between binding free energy and  extended  features.   Obviously, the determination of  $f_{\rm binding}$ is a major task of the present work.
% and will be discussed in more detail in Section   \ref{SolvationPrediction}.

\paragraph{Similarity assumption}

In the FFT,  we assume that molecules with similar microscopic features have similar macroscopic features, or physical observables. In the present application, we assume that protein complexes with similar microscopic features will have similar binding  free energies.  This assumption provides the basis for utilizing   supervised  machine learning algorithms to rank protein-ligand complexes.

In our earlier HPK model, we assume that molecules with similar features have the same set of parameters in a physical model. As a result, solvation or binding free energies are still computed based on a physical model, while a machine learning algorithm is used to find out the nearest neighbors for modeling  physical parameters.  In the present FFT, the binding free energy is not modeled by a physical model directly. However, the microscopic features are constructed from physical models.

%@@@@@@@@@@@@@@@

\subsection{Microscopic features }\label{features}

In physical models, such as MM PBSA and MM GBSA, the protein ligand binding affinity is given by  the combination of molecular mechanics  energy, solvation free energy, and entropy term
\begin{equation}
\label{BindingEnergy-LTR}
\Delta G=\Delta E_{\rm MM}+\Delta G_{\rm solv}-T \Delta S,
\end{equation}
where $\Delta E_{\rm MM}$, $\Delta G_{\rm solv}$, and $T  \Delta S$ are the molecular mechanics energy, solvation free energy, and entropy terms, respectively. Further, the molecular mechanics energy can be decomposed as  $E_{\rm Covalent}$,  which is the sum of bond, angle, and torsion energy terms, and $E_{\rm Noncovalent}$, which includes the van der Waals term and a Coulombic term $E_{\rm Coul}$ \cite{Greenidge:2012}. Equation (\ref{BindingEnergy-LTR}) is used as a guidance for the feature selection in our FFT-Score model.

\paragraph{Reaction field features}
Molecular electrostatics is of fundamental importance in the protein  solvation and binding processes \cite{Sharp:1990a, Honig:1995a,Gilson:1993}. In this work, we use a classical implicit solvent model, the PB theory, for modeling the molecular electrostatics in the solvent environment. This model is used for two purposes. On the one hand, the solvation effects during the protein ligand binding will be modeled via this theory. On the other hand, the electrostatic contribution to the protein ligand binding affinity is computed based on this model, as well.

For simplicity, we consider the linearized PB model in the pure water solvent, which is formulated as the following elliptic interface problem in mathematical terminology.  The governing equation is given by
\begin{equation}
\label{PBELTR}
-\nabla\cdot(\epsilon(\mathbf{r})\nabla\phi(\mathbf{r}))=\sum_{i=1}^{N_m} Q_i\delta(\mathbf{r}-\mathbf{r}_i),
\end{equation}
with the interface conditions
\begin{equation}
\label{Continuity-potentialLTR}
[\phi]|_\Gamma=0,
\end{equation}
and
\begin{equation}
\label{Continuity-fluxLTR}
[\epsilon\phi_{\mathbf{n}}]|_\Gamma=0,
\end{equation}
where $\phi$ is the electrostatics potential over the whole solvent solute domain, $Q_i$ is the partial charge located at $\mathbf{r}_i$ and $\delta(\mathbf{r}-\mathbf{r}_i)$ is the delta function at point $\mathbf{r}_i$. The permittivity function $\epsilon(\mathbf{r})$ is given by
\begin{equation}
\label{Permittivity-LTR}
\epsilon(\mathbf{r})=\left\{
                       \begin{array}{ll}
                         \epsilon_{\rm m}=1, & \ \mathbf{r}\in\Omega^{\rm m} \\
                         \epsilon_{\rm s}=80, & \ \mathbf{r}\in \Omega^{\rm s}
                       \end{array}
                     \right.
\end{equation}
where $\Omega^{\rm m}$ and $\Omega^{\rm s}$ are solute and solvent domains, respectively.  The two domains are separated by the molecular surface $\Gamma$.

The following Debye-Huckel type of boundary condition is imposed to make the PB model well posed
\begin{equation}
\label{DHBC-LTR}
\phi(\mathbf{r})=\sum_{i=1}^{N_m} \frac{Q_i}{4\pi \epsilon_{\rm s} |\mathbf{r}-\mathbf{r}_i|}, \ \mbox{if}\ \mathbf{r}\in\partial \Omega,
\end{equation}
where $\Omega=\Omega^{\rm m}\bigcup \Omega^{\rm s}$.

Molecular reaction field energy is computed by the following formula
\begin{equation}
\label{RFEnergy-LTR}
\Delta G_{\rm RF}= \sum_{i=1}^{N_m} \Delta G_{{\rm RF}i}
\end{equation}
where   the   $i$th atomic reaction field energy $\Delta G_{{\rm RF}i}$ is given by
\begin{equation}
\label{RFEnergy-atom}
  \Delta G_{{\rm RF}i} = \frac{1}{2}Q_i(\phi(\mathbf{r}_i) - \phi_{\rm h}(\mathbf{r}_i))
\end{equation}
where  $\phi_{\rm h}$ is obtained through solving the PB model with $\epsilon({\bf r})=1$ in the whole computational domain $\Omega$.
%Indeed, when $\epsilon_{\rm s}$ is set to be 1, the PB model in the pure water solvent is degenerated to the classical Poisson model without interface.
Note that atomic reaction field energies $\Delta G_{{\rm RF}i}$ are used as features in our FFT based solvation  model.

Here the reaction field energy gives a good description of the solvation free energy.
In our earlier study on the  solvation model, we found that reaction field energy related molecular descriptor provides a very accurate characterization of the solvation effects. The study of a large amount of small solute molecules demonstrates that by using these microscopic features  in the solvation model, the predicted solvation free energy is in an excellent agreement with the experimental solvation free energy.  For example,  the RMSE of our leave-one-out test for a large database of 668  molecules is around 1 kcal/mol \cite{BaoWang:2016LTR}.

Note that in Eq. (\ref{RFEnergy-LTR}), the whole reaction field energy is regarded as the sum of atomic reaction field energies. In the PB calculation, the solute molecule is usually assumed to be a homogeneous dielectric continuum with a uniform dielectric constant, which is an inappropriate assumption, since atoms in different environments should have different dielectric properties \cite{RenxiaoWang:2003CompareSF}. For this reason, we select the atomic reaction field energy as a microscopic feature and let the machine learning algorithm to automatically take care the possible difference in dielectric constants.
%This treatment is like a predictor corrector method, we first suppose a dielectric function to calculate the reaction field energy and then correct them according to the experimental results through using machine learning method.

\paragraph{Electrostatic binding features}
 By using the PB model, we can further obtain the electrostatics contribution to the protein-ligand binding affinity. The electrostatics binding free energy is calculated by
\begin{equation}
\label{Elec-binding-LTR}
\Delta G_{\rm el}=(\Delta G_{\rm RF})_{\rm Com}-(\Delta G_{\rm RF})_{\rm Pro}-(\Delta G_{\rm RF})_{\rm Lig}+\Delta G_{\rm Coul},
\end{equation}
where $\Delta G_{\rm el}$ is the electrostatics binding free energy between protein and ligand, $(\Delta G_{\rm RF})_{\rm Pro}$ and $(\Delta G_{\rm RF})_{\rm Lig}$ are the reaction field energies of the protein and ligand, respectively. Here  $\Delta G_{\rm Coul}$ is the Coulombic interaction between the two parts in the vacuum environment, which is computed as
\begin{equation}
\label{Coulomb-LTR}
\Delta G_{\rm Coul}=\sum_{i, j}\frac{Q_iQ_j}{r_{ij}},
\end{equation}
where $r_{ij}$ is the distance between two specific charges, and   indexes $i$ and $j$ run over all the atoms in the protein and ligand molecules, respectively. {  It is worthy to remind that the electrostatics binding free energy $\Delta G_{\text{el}}$ is a microscopic feature representing the contributions of solvation and Coulombic to the macroscopic binding free energy $\Delta G$.}
The PB model is solved by our in-house software, MIBPB \cite{Zhou:2006c,Yu:2007,Geng:2007a,DuanChen:2011a}, which is shown to be grid size independent. Its relative ranking orders of reaction field energy and binding free energy calculated  with  different grid sizes are consistent \cite{DDNguyen:2017a}. This numerical accuracy guarantees the preserving of relative ranking orders, which in turn  avoids  the influence on the prediction from numerical errors.

%PB accurate solver, relative ranking order preserving scheme, to avoid numerical error's influence on the model accuracy

%Briefly talk about PB theory and atomic reaction field energy, electrostatics binding affinity

%Electrostatics contribution to the binding

%In the ranking electrostatics can be used to model the solvation effects, which is tested on a large variety of small molecules
\paragraph{Atomic Coulombic interaction}
Coulombic energy plays an important role in the molecular mechanics energy \cite{massova2000combined, Kollman:2000, Greenidge:2012}. Coulombic energy calculation also depends on the dielectric medium. To this end, we considered the atomic Coulombic interactions  in vacuum environment. Specifically,  for the $i$th  atom in the protein molecule, we select the microscopic feature from atomic Coulombic energy as
\begin{equation}
\label{CoulombDescriptor-LTR}
(\Delta G_{{\rm Coul}})_i=\sum_j \frac{Q_iQ_j}{r_{ij}},
\end{equation}
where  the summation index $j$ runs over all the atoms in the ligand molecule. The Coulombic energy associated with the atoms in the ligand molecules can be defined analogously.

\paragraph{Atomic van der Waals interaction}

It was shown that van der Waals interactions play an important role in solvation analysis \cite{Gallicchio:2004,choudhury2005mechanism,Wagoner:2006,ZhanChen:2012,BaoWang:2015a}. We expect that van der Waals interactions are essential to binding process as well. In this work, we consider the 6-12 Lennard Jones (LJ)  interaction potential for modeling the van der Waals interactions
\begin{equation}
\label{LJPotential-LTR}
u_{ij}(\mathbf{r}_i, \mathbf{r}_j)=\epsilon_{ij}\left[\left(\frac{r_i+r_j}{||\mathbf{r}_i-\mathbf{r}_j||}\right)^{12}-2\left(\frac{r_i+r_j}{||\mathbf{r}_i-\mathbf{r}_j||}\right)^6 \right],
\end{equation}
where $r_i$ and $r_j$ are atomic radii of the $i$th and $j$th atoms, respectively. Here $\epsilon_{ij}$ measures the depth of the attractive well at $||\mathbf{r}_i-\mathbf{r}_j||=r_i+r_j$.
For features related to the van der Waals interactions, we select pairwise particles interactions as microscopic features for describing the van der Waals interactions between the protein and ligand. In these features,  each atom type is collected together,  and well-depth parameters $\epsilon_{ij}$ are left as training parameters in the subsequent ranking procedure.

\paragraph{Atomic solvent excluded surface area and molecular volume}

Molecular surface area and surface enclosed volume are usually  employed in scaled-particle theory (SPT)  to model the nonpolar solvation free energy \cite{Stillinger:1973,Pierotti:1976,Lum:1999} and/or  entropy contribution to the protein ligand binding affinity. In our FFT-BP, the solvent excluded surface  is employed for the conformation modeling of the solvated molecule. The molecular surface area associated with each atom type and molecular volume are used as microscopic features. These features are also computed by our in-house software, ESES \cite{ESES:2017}, in which a second order convergent scheme based on the level set theory and third order volume schemes are implemented. In ESES,  the molecular surface area is partitioned into atomic surface areas based on the power diagram theory.

\paragraph{Summary of microscopic features}
We consider microscopic features of a protein-ligand complex.  For the protein molecule, microscopic features are selected from  following types of atoms, namely, C, N, O, and S. For the ligand molecule, atomic features are collected from C, N, O, S, P, F, Cl, Br, and I. Here we drop features from hydrogen atoms (H) since  the positions of these atoms are not typically given in original X-ray crystallography data, and their information  may not be accurate. This selection of representative atoms is consistent with that of some other existing scoring functions, e.g., Cyscore \cite{Cao:2014Cyscore}, AutoDock Vina \cite{Trott:2010AutoDock}, and RF-Score \cite{Pedro:2010Binding}. In our model, we collect  { electrostatic binding free energy}, atomic reaction field energies, molecular reaction field energy, atomic van der Waals and Coulombic interactions, atomic surface areas, and molecular volume as the building block of feature space. Due to the fact that binding is a thermodynamic process, the change of the atomic reaction field energies, atomic surface areas, and molecular volumes between the bounded and unbounded states are selected as microscopic features as well.

For the atomic features  associated with each type of element,  we consider  their corresponding statistical quantities, i.e., maximum, minimum and average, as features. Similarly, maximum, minimum and average of   absolute values of atomic electrostatic features  are also used features. { All features used in the current work are summarized in Table \ref{tab:features}.}

{ 
	\begin{center}
		\begin{longtable}{p{10cm}p{7cm}}	
			\caption{List of features and software used in protein-ligand binding energy prediction. Atom types X selected for protein are C, N, O and S. Atom types X selected for complex and ligand ar C, N, O, S, P, F, Cl, Br and I. All structure inputs in each feature calculation are in PQR format. The procedure for acquiring this format is discussed in Section \ref{data_process}}\label{tab:features}\\
			\hline \multicolumn{1}{l}{Features} &\multicolumn{1}{l}{Software} \\ \hline
			\endfirsthead
			{{\bfseries \tablename\ \thetable{} -- \footnotesize{continued}}} \\
			\hline \multicolumn{1}{l}{Features} &\multicolumn{1}{l}{Software} \\ \hline
			\endhead
			\endfoot
			\hline
			\endlastfoot
			Reaction field Energies for complex/protein/ligand  & MIBPB\newline (\url{http://weilab.math.msu.edu/MIBPB/}\vspace*{1ex})\\
			Electrostatic binding free energies & MIBPB and Python \vspace*{1ex}\\ 
			Molecular volumes for complex/protein/ligand& ESES \newline(\url{http://weilab.math.msu.edu/ESES/})  \vspace*{1ex}\\
			Molecular surface areas for complex/protein/ligand & ESES  \vspace*{1ex}\\
			Atomic van der Waals interactions between X atoms in protein and Y atoms in ligand & Python \vspace*{1ex}\\
			Statistical quantities (sum,mean,max,min) of atomic reaction field energies for X atoms in complex/protein/ligand & MIBPB and Python \vspace*{1ex} \\
			Statistical quantities (sum,mean,max,min) of atomic reaction field energies for X atoms in altogether complex, protein and ligand & MIBPB and Python \vspace*{1ex} \\
			Statistical quantities (sum,mean,max,min) of atomic reaction field energies for all atoms in complex/protein/ligand & MIBPB and Python \vspace*{1ex} \\
			Statistical quantities (sum,mean,max,min) of atomic reaction field energies for all atoms in altogether  complex, protein and ligand & MIBPB and Python \vspace*{1ex} \\
			Statistical quantities (sum,mean,max,min) of atomic surface areas for X atoms in complex/protein/ligand & ESES and Python\vspace*{1ex}\\
			Statistical quantities (sum,mean,max,min) of atomic surface areas for X atoms in altogether complex, protein and ligand & ESES and Python\vspace*{1ex}\\
			Statistical quantities (sum,mean,max,min) of atomic surface areas for all atoms in complex/protein/ligand & ESES and Python\vspace*{1ex}\\
			Statistical quantities (sum,mean,max,min) of atomic surface areas for all atoms in altogether complex, protein and ligand & ESES and Python\vspace*{1ex}\\
			Statistical quantities (sum,mean,max,min) of atomic Coulombic energies for X atoms in complex/protein/ligand & Python \vspace*{1ex}\\
			Statistical quantities (sum,mean,max,min) of atomic Coulombic energies for X atoms in altogether complex, protein and ligand & Python \vspace*{1ex}\\
			Statistical quantities (sum,mean,max,min) of atomic Coulombic energies for all atoms in complex/protein/ligand & Python \vspace*{1ex}\\
			Statistical quantities (sum,mean,max,min) of atomic Coulombic energies for all atoms in altogether complex, protein and ligand & Python \vspace*{1ex}\\
			Statistical quantities (sum,mean,max,min) of atomic charges for X atoms in complex & Python \vspace*{1ex}\\
			Statistical quantities (sum,mean,max,min) of atomic charges for all atoms in complex & Python \vspace*{1ex}\\
			Statistical quantities (sum,mean,max,min) of absolute values of atomic charges for X atoms in complex & Python \vspace*{1ex}\\
			Statistical quantities (sum,mean,max,min) of absolute values of atomic charges for all atoms in complex & Python \vspace*{1ex}\\
			Volume change between bounded and unbounded states & ESES and Python\\
			Statistical quantities (sum,mean,max,min) of atomic reaction field energies change of X atoms in protein/ligand & MIBPB and Python \vspace*{1ex}\\
			Statistical quantities (sum,mean,max,min) of atomic reaction field energies change of X atoms in both protein and ligand & MIBPB and Python \vspace*{1ex}\\
			Statistical quantities (sum,mean,max,min) of atomic reaction field energies change of all atoms in protein/ligand & MIBPB and Python \vspace*{1ex}\\
			Statistical quantities (sum,mean,max,min) of atomic reaction field energies change of all atoms in both protein and ligand & MIBPB and Python \vspace*{1ex}\\
			Statistical quantities (sum,mean,max,min) of atomic area change of X atoms in protein/ligand& ESES and Python \vspace*{1ex}\\
			Statistical quantities (sum,mean,max,min) of atomic area change of X atoms in both protein and ligand& ESES and Python \vspace*{1ex}\\
			Statistical quantities (sum,mean,max,min) of atomic area change of all atoms in protein/ligand & ESES and Python \vspace*{1ex}\\
			Statistical quantities (sum,mean,max,min) of atomic area change of all atoms in both protein and ligand & ESES and Python \vspace*{1ex}\\
			
		\end{longtable}
	\end{center}
}

%List of features, statistical features

\subsection{Machine learning algorithm}
Many machine learning algorithms, including support vector machine, decision tree learning, random forest, and deep neural network can be employed.
A specific machine learning algorithms utilized in the present study to protein ligand binding affinity scoring is an MART algorithm.  MART is a list-wise LTR algorithm, for a given training set with feature vectors and associated ranking order (here we simply using the protein-ligand binding affinity as this label value), it trains a function that optimally simulates the relation between features and labels. When applied to a protein-ligand complex in the test set, this trained function acts on the corresponding features and gives a predicted value.  The predicted value reflects the binding affinity of the complex in the test set. In the web-search community, LambdaMART is one of the state-of-the-art LTR algorithms, here LambdaMART is a coupling of Lambda and MART. Compared to the classical MLR model for training functions that  link  features and labels, MART can capture the nonlinear relationship. Furthermore, compared to most neuron network based algorithms, it is more efficient. MART also named GBDT (gradient boosting decision tree) is a very efficient ensemble method for regression.  Meanwhile, due to the boosting of the weaker learners (usually quite simple models like decision tree), the over-fitting problem can be avoided effectively. The principles of the GBDT are summarized as following:
\begin{itemize}
\item For the training set, GBDT successively learns the weak learners, and each weak learner is a regression tree with quite a few levels for fitting the residual of the previous forest compared to the training set. This procedure starts from a regression tree for fitting the training set, and the regression tree is added into the forest gradually. Each succeed regression tree is used for fitting the residual of the previous forest.

\item Instead of counting the whole contribution from each regression tree, shrinkage is adopted, which is a weight of the regression tree. This weight is obtained through solving an optimization problem via the simple line searching algorithm.

\item Weighted contributions from the whole regression trees are presented in the final scoring function, which is the boosting of simple regression trees. Due to the simplicity of each regression tree, the over-fitting problem can be bypassed efficiently.
\end{itemize}

In summary, the MART  learns a function between features and the binding free energy through the training set. In the testing step, this function assigns a predicted binding affinity to each sample in the testing set, and the ranking position of a given sample is determined through the obtained score. This ranking method is significantly different from the classical pairwise approaches, e.g., RankSVM \cite{RankSVM:2002, RankSVM:2006, RankSVM:2014}, where ranking is based on the pairwise comparison between all sample pairs in the training set. A major drawback of these approaches is that they assumes the same penalty for all pairs. In contrast,   we only care about  a few top   ranking results for a given query in most applications. For more comprehensive and mathematical description of the MART, reader is   referred to the literature \cite{LambdaMart:2010, GBDT:2001}. Many other LTR algorithms can be used in our framework as well, like LambdaMART \cite{LambdaMart:2010, GBDT:2001}, ListNet \cite{ListNet:2007}, etc.

\subsection{Method for binding affinity prediction  }\label{BindingPrediction}

In this subsection, we discuss the FFT prediction of the binding  free energy  of a given target protein-ligand complex AB. Based on our assumption that binding free energy is a functional of feature vectors,  we construct a feature function around the target molecular complex and use it to predict the binding  free energy.  Even though the exact form of the function between feature and binding affinity is unknown, locally it can be approximated by a linear function. In other words, locally we assume the binding affinity is a linear function of  microscopic feature vectors.

The importance of various features can be ranked automatically during the machine learning procedure, and thus the number of influential features ($n$) can be reduced by selecting features of  top importance to represent the binding affinity. We assume that  target molecular complex AB is characterized by its feature vector $\mathbf{x}_{\rm AB}=({\bf x}_{{\rm AB}1}, {\bf x}_{{\rm AB}2}, \cdots, {\bf x}_{{\rm AB}n})$, where $n$ is the dimension of the microscopic feature space, i.e., the space of all microscopic feature vectors. We also assume that by using the LTR algorithm,  we can find   top $m$ nearest neighbors from the training set. The  extended feature  vectors of these nearest neighbor complexes are given by  $\{ {\bf v}_i=(\mathbf{x}_i,  \Delta G_i)\}_{i=1}^m$. In general,  the dimension of the feature space is  much larger than  the number of nearest neighbors used, i.e., $m \ll n$. Therefore, the direct least square approach may lead to over-fitting.  To avoid over-fitting, we utilize a Tikhonov regularization based least square algorithm  for training the binding affinity function. From the extended feature vectors, we can set up the following set of   equations
\begin{equation}
\label{Regression}
\left(
  \begin{array}{c}
    \Delta    G_1 \\
    \Delta   G_2 \\
    \vdots \\
    \Delta   G_m \\
  \end{array}
\right)=
\left(
  \begin{array}{cccc}
    x_{11} & x_{12} & \cdots & x_{1n} \\
    x_{21} & x_{22} & \cdots & x_{2n} \\
    \vdots & \vdots & \vdots & \vdots \\
    x_{m1} & x_{m2} & \cdots & x_{mn} \\
  \end{array}
\right)\left(
         \begin{array}{c}
           w_1 \\
           w_2 \\
           \vdots \\
           w_n \\
         \end{array}
       \right)+
       \left(
         \begin{array}{c}
           b \\
           b \\
           \vdots \\
           b \\
         \end{array}
       \right),
\end{equation}
where $w_i=w_i({\bf v}_1,  {\bf v}_2, \cdots  {\bf v}_m) $ and $b=b({\bf v}_1,  {\bf v}_2, \cdots  {\bf v}_m) $ define the function for $\Delta G_i$. By the similarity assumption, the same functional form can be used for target complex AB. For further derivation, we rewrite Eq. (\ref{Regression}) as
\begin{equation}
\label{Regression2}
\Delta   \mathbf{G}={\bf x}\mathbf{w}+b\mathbf{1},
\end{equation}
where $\Delta  \mathbf{G}=\left(\Delta   G_1, \Delta   G_2, \cdots, \Delta   G_m\right)^T$, $\mathbf{w}=\left(w_1, w_2, \cdots, w_n\right)^T$, $\mathbf{1}$ is an $m$-dimensional column vector with all elements equaling 1, and matrix ${\bf x}$ is given by
$$
{\bf x}=\left(
  \begin{array}{cccc}
    x_{11} & x_{12} & \cdots & x_{1n} \\
    x_{21} & x_{22} & \cdots & x_{2n} \\
    \vdots & \vdots & \vdots & \vdots \\
    x_{m1} & x_{m2} & \cdots & x_{mn} \\
  \end{array}
\right).
$$

To avoid over-fitting, we add an $L_2$ penalty to the weight vector $\mathbf{w}$, and solve  Eq. (\ref{Regression2}) as an optimization problem
\begin{equation}
\label{Optimization}
\min_{\mathbf{w}, b} ||\Delta   \mathbf{G}-{\bf x} \mathbf{w}-b\mathbf{1}||_2^2+\lambda||\mathbf{w}||_2^2:=\min_{\mathbf{w}, b} {\bf F},
\end{equation}
where $\lambda$ is a regularization parameter and is set  to 10 in this work, and $||*||_2$ denotes the $L_2$ norm of the quantity $*$.

By solving $\frac{\partial {\bf F}}{\partial \mathbf{w}}=0$, we have
\begin{equation}
\label{weightssolution}
\mathbf{w}=\left({\bf x}^T{\bf x}+\lambda \mathbf{I}\right)^{-1}\left({\bf x}^T\Delta  \mathbf{G}-{\bf x}^T(b\mathbf{1})\right),
\end{equation}
where $I$ is an $m\times m$ identity matrix.

To determine $b$ from Eq. (\ref{Optimization}), we relax $b\mathbf{1}$ to an arbitrary vector such that $\mathbf{b}=\left(b_1, b_2, \cdots, b_m\right)^T$. By solving $\frac{\partial {\bf F}}{\partial \mathbf{b}}=0$, we have
\begin{equation}
\label{interceptionsolution1}
\mathbf{b}=\Delta   \mathbf{G}-{\bf x}\mathbf{w}.
\end{equation}
An unbiased estimation of $b$ is given by
\begin{equation}
\label{interceptionsolution2}
b=\frac{\sum_{i=1}^m(\Delta   \mathbf{G}-{\bf x}\mathbf{w})_i}{m},
\end{equation}
where $(\Delta   \mathbf{G}-{\bf x}\mathbf{w})_i$ is the $i$th component of the vector $\Delta  \mathbf{G}-\mathbf{xw}$.

The optimization problem in Eq. (\ref{Optimization}) is solved by alternately  iterating Eqs. (\ref{weightssolution}) and  (\ref{interceptionsolution2}), which is essentially an expectation - maximization (EM) algorithm.
%We summarize the algorithm for solving Eq. (\ref{Optimization}) in Algorithm \ref{Alg:EM}.
%
%\begin{algorithm}
%\caption{EM algorithm for solving the optimization problem Eq. (\ref{Optimization})}\label{Alg:EM}
%\begin{algorithmic}[1]
%%\Procedure{}{}
%\State \textbf{Initialize:} $\mathbf{w}=\mathbf{0}$, $b=\frac{\sum_{i=1}^m \Delta \mathbf{G}}{m}$, $F_1=||\Delta \mathbf{G}-\mathbf{xw}-b\mathbf{1}||_2^2+\lambda||\mathbf{w}||_2^2$, $F_2=F_1+100$.
%\State \textbf{do while} ($|F_1-F_2|>\epsilon_0$)
%\State \hskip 1.0cm Update $F_2$: $F_2\leftarrow F_1$.
%\State \hskip 1.0cm Update $\mathbf{w}$: $\mathbf{w}\leftarrow \left({\bf x}^T{\bf x}+\lambda \mathbf{I}\right)^{-1}\left({\bf x}^T\Delta \mathbf{G}-{\bf x}^T(b\mathbf{1})\right)$.
%\State \hskip 1.0cm Update $b$: $b\leftarrow \frac{\sum_{i=1}^m(\Delta \mathbf{G}-{\bf x}\mathbf{w})_i}{m}$.
%\State \hskip 1.0cm Update $F_1$: $F_1\leftarrow ||\Delta \mathbf{G}-{\bf x}\mathbf{w}-b\mathbf{1}||_2^2+\lambda||\mathbf{w}||_2^2$.
%\State \textbf{enddo}
%%\EndProcedure
%\end{algorithmic}
%\end{algorithm}
%In Algorithm \ref{Alg:EM}, $\epsilon_0$ is the threshold parameter for control the convergence of the solution to the optimization problem and is set to  0.01 in this work.

After  obtaining optimized weights $\mathbf{w}$ for the feature vector ${\bf x}$ and hyperplane height $b$, the binding free energy of   target molecular complex  AB can be  predicted as
\begin{equation}
\label{solvationenergy}
\Delta     G_{\rm AB}=b+\sum_{i=1}^n w_i{\bf x}_{{\rm AB}i}.
\end{equation}
Equation (\ref{solvationenergy}) can be regarded as a linear  approximation of the binding free energy functional $\Delta   G_{\rm AB}=f(\mathbf{x}_{\rm AB},{\bf v}_1,  {\bf v}_2, \cdots  {\bf v}_m)$.

Alternatively, we can also directly obtain the binding affinity of the target complex AB from the LTR ranking value if the ranking algorithm attempts to fit the target value. For general LTR algorithms, especially pairwise ranking algorithms, the direct use of the ranking score as a predicted binding affinity is not appropriate. However, the proposed protocol also applies to this scenario. These two approaches are compared in this present work.

\section{Numerical results} \label{Results}

In this section, we explore the validity, demonstrate the performance, and examine the limitation of the proposed FFT-BP. First, we describe datasets used in this work.
 Then, we examine  whether FFT-BP's performance depends on  protein clusters, where each cluster contains one specific protein and tens or hundreds of ligands. Our test
 on a validation set of 1322 protein-ligand complexes from 7 clusters indicates that the performance of the proposed FFT-BP does not depend on protein clusters. By using the same test set, we also study the impact of cut-off distance to FFT-BP prediction. Here cut-off distance refers to a protein feature evaluation truncation distance. Protein atoms within the cut-off distance are allowed to contribute the atomic feature selection and calculation (except for molecule-wise features, such as volume, electrostatic solvation free energy, electrostatic binding free energy, etc). To further benchmark  the accuracy of the present FFT-BP, we carry out a five-fold cross validation on training set ($N=3589$), which is derived from the PDBBind v2015 refined set \cite{PDBBind:2015}. Finally, we provide  blind predictions on a benchmark set of 100 protein-ligand complexes \cite{RenxiaoWang:2003CompareSF}, the PDBBind v2007 core set ($N=195$) \cite{Pedro:2010Binding}, and the PDBBind v2015 core set ($N=195$) \cite{PDBBind:2015}.

\subsection{Dataset preparation}
All data sets used in the present work are obtained from the PDBBind database \cite{PDBBind:2015}, in which  the PDBBind v2015 refined set of 3,706 entries was  selected from a general set of 14,620 protein-ligand complexes with good quality, filtered over  binding data, crystal structures, as well as the nature of the complexes  \cite{PDBBind:2015}. Due to the feature extraction, a pre-processing of data is required in the present method.

\subsubsection{Datasets}

This work utilizes one  validation set ($N=1322$), one training set ($N=3589$), and three test sets ($N=195, N=195$ and $N=100$) as described below.

\paragraph{Validation set ($N=1322$)}
To explore the cluster dependence (or independence)  and  the optimal cut-off distance of the present FFT-BP, we select a subset of the PDBBind v2015 refined set with 1322 complexes in 7 different clusters. Each cluster contains one protein and a large number,  ranging from 93 to 333, of small ligand molecules. With this validation, we examine whether the predictions inside various clusters are more accurate than the overall prediction regardless of clusters. The performance dependence of the cut-off distance is also explored with this set.

\paragraph{Training sets}
For the PDBBind v2015 refined set, we carry our FFT  microscopic feature extraction via appropriate force field parametrization described below, which leads to a parametrized set of 3589 protein-ligand complexes.   The training set is employed to train our FFT model according to each test. Whenever a test set is studied, its entries are carefully excluded from the training set of 3589 complexes and then, the model is trained without any test set molecule. { Similarly, we apply our FFT approach for training another training set, namely PDBBind v2007 refined set, comprising 1082 complexes.}

\paragraph{Test sets}
Three test sets are standard ones described in the literature. PDB IDs of the training set and the validation set are given in the Supporting material.

{\it The PDBBind v2015 core set} of 195   benchmark-quality complexes is employed as a test set. According to the literature \cite{PDBBind:2015},  the PDBBind v2015 core set was selected  with an emphasis on the diversity in structures and binding data. It contains 65 representative clusters from the PDBBind v2015 refined set. For each cluster, it must have  at least five protein-ligand complexes and three complexes,   one with the highest binding constant,  another with the lowest binding constant, and the other with a medium binding constant were selected for the PDBBind v2015 core set \cite{PDBBind:2015}.
	
We also consider two additional test sets, {\it the PDBBind v2007 core set} of 195 complexes  \cite{RenxiaoWang:2009Compare} and {\it the  benchmark set} of 100 complexes \cite{RenxiaoWang:2003CompareSF} to benchmark the proposed FFT-BP against a large number of scoring functions.

%When the training set ($N=3589$) is applied to a test set, we first exclude all the overlapping entries between the training and the given test and re-train the training set for the specific test.

\subsubsection{Data pre-processing}\label{data_process}
%Binding energy formula, to kcal/mol transformation
FFT-BP utilizes microscopic features, which requires appropriate feature extraction from the data set.  Before the feature generation,  structure optimization and  force field assignment are carried out. Protein structures with corresponding ligand are prepared with the protein preparation wizard utility of the Schr\"{o}dinger 2015-2 Suite \cite{Friesner:2004Schrodinger, Madhavi:2013Schrodinger} with default parameters except filling the missing side chains. The protonation states for ligands are generated using Epik state penalties and the H-bond networks for the complex are further optimized using PROPKA at pH 7.0 \cite{Michal:2011Schrodinger, Olsson:2011Schrodinger}. The restrained minimization on heavy atoms for the complex structures are finally performed with OPLS 2005 force field  \cite{OPLS:1988}. The atomic radii and charges for the complexes are parametrized by Amber tool14 \cite{AMBER15}. For ligand molecules, charges are calculated by the antechamber module with AM1-BCC semi-empirical charge method and the atomic radii are assigned by using the mbondi2 radii set \cite{Jakalian:2002}. For protein molecules, radii and charges of each atom are parametrized by the Amber ff14SB general force field with  tleap module \cite{AMBER15}.

Protein features are extracted with a cut-off distance. Specifically, we first find a   tight bounding   box  containing the ligand, then extend feature generation domain  along  all directions around the box to a cut-off distance.   We provide all the data involved in this work in the Supporting material, in which some protein-ligand structures that need specific treatments are highlighted.

In the PDBBind database, the protein ligand binding affinity is provided in term of $pK_d$.
 We   convert all the energy unit in the PDBBind database to kcal/mol.  To derive the unit convert formula, one notes that
$$
\Delta G=RT\ln k_d=-RT\ln K_{eq},
$$
where $\Delta G$ is the Gibbs free energy, $k_d$ is the disassociation constant, and  $R$ is the gas constant. Since $pK_d=-\log_{10} K_d$, then at the room temperature, $T=298.15 K$, one has the following relation between these two units
\begin{equation}
\label{UnitConversion-LTR}
\Delta G=-1.3633\  pK_d.
\end{equation}
%Add protocol here, BW to be done

\subsection{Validation}

In this section, we explore the properties of FFT-BP and validate its performance.   The following two important issues are  examined in several existing scoring functions. The first issue is related to the protein-ligand binding affinity prediction of diverse multiple clusters,   especially  clusters with limited experimental data. Another issue is that a scoring method should be optimized with a cut-off distance in the feature extraction  to maintain sufficient accuracy and avoid unnecessary  feature calculations. In the existing work, the LTR based scoring functions can predict  cross-cluster binding affinity  well \cite{WeiZhang:2015}. For the random forest and some other machine learning algorithms, one typically selects a cut-off distance of  12 \AA, in the  protein feature  calculation \cite{Ballester:2014}.

In this work,   we  demonstrate the capability of the FFT-BP for the accurate cross-cluster binding affinity prediction. Additionally,  we explore the optimal  cut-off distance for   FFT-BP feature extraction. Finally, since the accuracy of the FFT-BP predictions depends on the numbers of the nearest neighbors and top features,  we investigate robustness of the proposed FFT-BP with respect to choices of  the nearest neighbors and top features.

Two sets of protein-ligand complexes, i.e., the validation set ($N=1322$) and the training set $(N=3589)$, are employed in this validation study.

\subsubsection{Validation on the validation set ($N=1322$)}

%searching part, more chemical information about the complexes should be incorporated. Ranking among the whole data do not completely consistent with our basic assumption. Partitioning the training data into several queries with each group complexes share some common properties is under our consideration.

\begin{table}
\centering
\caption{The RMSEs (kcal/mol) for the five-fold validation on the 7 clusters of the validation set and on the whole  validation set ($N=1322$) with 10 different cut-off distances in the feature extraction.}
\begin{tabular}{ c|l|cccccccccc}
\hline
\multirow{2}{*}{Test set}&\multirow{2}{*}{Group }&\multicolumn{10}{c }{Cut-off distance}\\
%\hline
               &        &5 \AA   &10 \AA   &15 \AA    &20 \AA  &25 \AA   &30 \AA  &35 \AA  &40 \AA  &45 \AA  &50 \AA\\ \hline
\multirow{6}{*}{Cluster 1} &Group1     &1.90&  1.86&  1.76&  1.73&  1.81&  1.90&  1.81&  1.82&  1.82&  1.86\\
                        &Group2       &2.07&  2.15&  2.38&  2.23&  2.35&  2.25&  2.21&  2.24&  2.24&  2.24\\
                        &Group3       &2.31&  1.98&  2.04&  1.95&  1.85&  1.87&  1.87&  1.89&  1.90&  1.90\\
                        &Group4       &1.89&  1.75&  1.58&  1.63&  1.63&  1.67&  1.62&  1.66&  1.65&  1.65\\
                        &Group5       &2.35&  2.22&  2.09&  2.05&  2.14&  1.67&  2.10&  2.12&  2.12&  2.13\\
                        &Average      &2.11&  2.01&  1.99&  1.93&  1.97&  2.13&  1.93&  1.96&  1.96&  1.97\\
\hline
\multirow{6}{*}{Cluster 2} &Group1     &1.39&  1.33&  1.31&  1.32&  1.39&  1.43&  1.46&  1.42&  1.42&  1.44\\
                        &Group2       &1.66&  1.24&  1.31&  1.23&  1.19&  1.14&  1.15&  1.19&  1.19&  1.19\\
                        &Group3       &1.39&  1.28&  1.14&  1.21&  1.28&  1.31&  1.37&  1.37&  1.32&  1.32\\
                        &Group4       &1.44&  1.33&  1.35&  1.36&  1.37&  1.38&  1.35&  1.40&  1.30&  1.40\\
                        &Group5       &1.53&  1.44&  1.38&  1.49&  1.36&  1.37&  1.38&  1.33&  1.38&  1.29\\
                        &Average      &1.49&  1.33&  1.34&  1.32&  1.32&  1.33&  1.34&  1.35&  1.33&  1.33\\
\hline
\multirow{6}{*}{Cluster 3} &Group1     &2.56&  2.40&  2.65&  2.41&  2.53&  2.62&  2.61&  2.60&  2.60&  2.60\\
                        &Group2       &2.07&  2.13&  2.08&  2.10&  2.11&  2.11&  2.11&  2.09&  2.09&  2.09\\
                        &Group3       &1.54&  1.53&  1.52&  1.57&  1.55&  1.51&  1.52&  1.50&  1.50&  1.50\\
                        &Group4       &1.82&  1.75&  1.70&  1.71&  1.64&  1.68&  1.70&  1.72&  1.72&  1.72 \\
                        &Group5       &2.14&  2.23&  2.20&  2.15&  2.18&  2.26&  2.26&  2.27&  2.27&  2.27\\
                        &Average      &2.05&  2.03&  2.07&  2.01&  2.03&  2.08&  2.08&  2.08&  2.08&  2.08\\
\hline
\multirow{6}{*}{Cluster 4} &Group1     &1.59&  1.78&  1.80&  1.88&  1.76&  1.68&  1.72&  1.72&  1.76&  1.81\\
                        &Group2       &1.41&  1.47&  1.53&  1.25&  1.34&  1.39&  1.37&  1.34&  1.37&  1.37\\
                        &Group3       &1.58&  1.46&  1.50&  1.59&  1.56&  1.52&  1.55&  1.55&  1.53&  1.55\\
                        &Group4       &1.91&  1.76&  1.76&  1.87&  1.83&  1.84&  1.80&  1.78&  1.81&  1.87\\
                        &Group5       &1.57&  1.54&  1.61&  1.73&  1.81&  1.84&  1.74&  1.67&  1.75&  1.74\\
                        &Average      &1.62&  1.61&  1.64&  1.68&  1.67&  1.67&  1.65&  1.62&  1.65&  1.68\\
\hline
\multirow{6}{*}{Cluster 5} &Group1     &2.01&  2.43&  1.83&  1.64&  1.60&  1.65&  1.67&  1.69&  1.65&  1.67\\
                        &Group2       &2.15&  2.08&  1.89&  1.88&  1.92&  1.86&  1.94&  1.88&  1.92&  1.87\\
                        &Group3       &2.52&  2.26&  2.54&  2.42&  2.41&  2.37&  2.39&  2.40&  2.35&  2.35\\
                        &Group4       &1.65&  1.70&  1.30&  1.37&  1.25&  1.33&  1.35&  1.36&  1.34&  1.30\\
                        &Group5       &3.18&  2.87&  2.89&  2.49&  2.59&  2.54&  2.56&  2.67&  2.65&  2.65\\
                        &Average      &2.39&  2.31&  2.18&  2.03&  2.03&  2.02&  2.05&  2.07&  2.06&  2.05\\
\hline
\multirow{6}{*}{Cluster 6} &Group1     &3.17&  2.99&  3.03&  2.95&  3.00&  3.02&  2.90&  2.92&  2.92&  2.92\\
                        &Group2       &2.09&  1.83&  1.83&  1.82&  1.91&  1.83&  1.88&  1.86&  1.86&  1.86\\
                        &Group3       &1.68&  1.71&  1.55&  1.65&  1.69&  1.55&  1.63&  1.58&  1.56&  1.56\\
                        &Group4       &1.73&  1.69&  1.55&  1.60&  1.60&  1.51&  1.58&  1.58&  1.59&  1.59\\
                        &Group5       &2.30&  1.97&  2.04&  2.13&  2.03&  2.06&  2.05&  2.05&  2.05&  2.05\\
                        &Average      &2.26&  2.09&  2.08&  2.09&  2.10&  2.07&  2.06&  2.06&  2.06&  2.06\\
\hline
\multirow{6}{*}{Cluster 7} &Group1     &1.83&  1.97&  2.16&  1.93&  1.68&  1.66&  2.01&  1.90&  1.88&  1.91\\
                        &Group2       &1.92&  1.99&  1.97&  1.97&  2.00&  1.93&  2.09&  2.06&  2.28&  2.21\\
                        &Group3       &1.68&  1.69&  1.45&  1.39&  1.35&  1.39&  1.44&  1.51&  1.54&  1.45\\
                        &Group4       &2.27&  2.11&  2.13&  1.91&  2.14&  2.14&  2.39&  2.36&  2.27&  2.42\\
                        &Group5       &1.76&  1.40&  1.29&  1.32&  1.41&  1.35&  1.38&  1.39&  1.44&  1.45\\
                        &Average      &1.90&  1.83&  1.81&  1.71&  1.73&  1.71&  1.88&  1.86&  1.90&  1.91\\
\hline
Average over all 7 clusters&             &1.90&  1.88&  1.87&  1.83&  1.84&  1.84&  1.85&  1.85&  1.85&  1.86\\
\hline \hline
\multirow{6}{*}{Whole set}  &Group1       &1.81&  1.55&  1.62&  1.57&  1.67&  1.69&  1.66&  1.55&  1.52&  1.57\\
                        &Group2       &1.63&  1.76&  1.62&  1.69&  1.64&  1.67&  1.55&  1.71&  1.73&  1.69\\
                        &Group3       &1.71&  1.58&  1.65&  1.65&  1.65&  1.55&  1.55&  1.63&  1.51&  1.66\\
                        &Group4       &1.73&  1.62&  1.65&  1.57&  1.56&  1.53&  1.78&  1.57&  1.69&  1.59\\
                        &Group5       &1.64&  1.65&  1.59&  1.64&  1.65&  1.68&  1.60&  1.63&  1.67&  1.64\\
                        &Average      &1.70&  1.63&  1.63&  1.63&  1.64&  1.63&  1.64&  1.63&  1.64&  1.64\\
\hline
\end{tabular}
\label{RMSerrorlist1322Test}
\end{table}

\begin{figure}[!ht]
\small
\centering
\includegraphics[width=6.5cm,height=5.5cm]{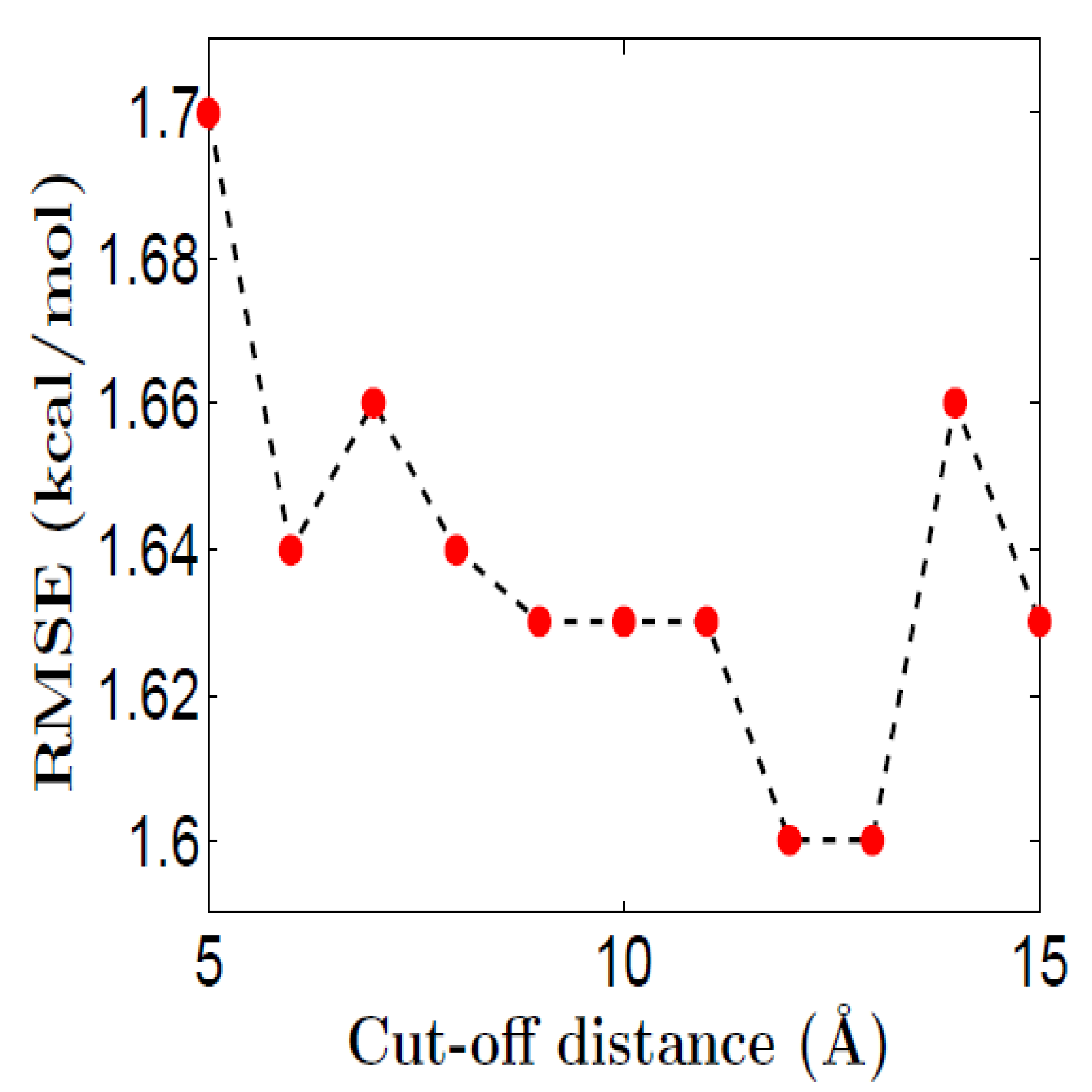}%{CutoffDistance.png}
\caption{The prediction RMSE vs the cut-off distance for the validation set ($N=1322$).}
\label{Cutoff}
\end{figure}
We validate the proposed FFT-BP on the validation set of 1322 complexes. We utilize the five-fold cross validation strategy to test the model and determine optimal cut-off distance. In this strategy, the validation set of 1322 complexes is     randomly partitioned into five essentially equal sized subsets. Of the five subsets, a single subset is retained as the test set for testing the FFT-BP, and the remaining four subsets are used as training data.  First, we run a coarse test with cut-off distance from 5 to 50 \AA \  using 5 \AA\ as the step size, which helps to determine a rough optimal cut-off distance. Second, we carry a refined search for the optimal the cut-off distance based on coarse test results with a step of size 1 \AA. At a given cut-off size, we do the five-fold cross validation on the validation set of 1322 complexes, together with the five-fold cross validation on each of 7 clusters. Table \ref{RMSerrorlist1322Test} lists the RMSEs on all the five-fold cross validation with cut-off distance 5 to 50 \AA \ and step size 5 \AA.

%Features can model the global binding affinity function
%Target independent

\begin{table}
\centering
\caption{The RMSEs  (kcal/mol)  for the five-fold test on the   validation set  ($N=1322$) with FFT-BP  calculated at different cut off distances. }
\begin{tabular}{c|ccccccccccc}
\hline
\multirow{2}{*}{Group } &\multicolumn{10}{c }{Cut-off distance}\\
      &5 \AA&6 \AA   &7 \AA   &8 \AA  &9 \AA   &10 \AA  &11 \AA  &12 \AA  &13 \AA & 14 \AA &15 \AA\\
\hline
Group1       &1.81&   1.68&   1.80&  1.58&   1.61&  1.55&  1.49&  1.62& 1.50& 1.60& 1.62\\
Group2       &1.63&   1.67&   1.61&  1.79&   1.67&  1.76&  1.63&  1.65& 1.76& 1.72& 1.62\\
Group3       &1.71&   1.68&   1.57&  1.65&   1.61&  1.58&  1.80&  1.62& 1.56& 1.71& 1.65\\
Group4       &1.73&   1.56&   1.46&  1.58&   1.64&  1.62&  1.74&  1.58& 1.55& 1.70& 1.65\\
Group5       &1.64&   1.57&   1.82&  1.57&   1.60&  1.65&  1.46&  1.56& 1.59& 1.59& 1.59\\
Average      &1.70&   1.64&   1.66&  1.64&   1.63&  1.63&  1.63&  1.60& 1.60& 1.66& 1.63\\
\hline
\end{tabular}
\label{RMSerrorlist1322Test-fine}
\end{table}

Results in Table \ref{RMSerrorlist1322Test} indicate that: 1) Overall, prediction over the whole set of 1322 complexes gives better results than predictions on individual  clusters. Therefore, the proposed method favors  blind cross-cluster predictions. 2) According the results from the whole validation set tests, feature cut-off distance at 10 \AA\  has reached an optimal value. This distance is actually consistent with the explicit solvent modeling in which a 10 \AA\ cut-off
distance is designed to account for long range electrostatic interactions.  To better estimate  the optimal cut-off distance, we carry out a more accurate searching in the range of 5 to 15 \AA\ distance with a step size of 1 \AA.  Table \ref{RMSerrorlist1322Test-fine} lists the RMSEs of the five-fold cross validation on the whole validation set of 1322 complexes.   These results show that 12 \AA\ is the optimal cut-off distance in the searched solution space, which is consistent with that used in the RF-Score \cite{Ballester:2014}. We plot the relation between the cut-off distance and prediction error in Fig. \ref{Cutoff}.  In the rest of this work, the cut-off distance of 12 \AA\ is utilized.

%At short range to the ligand molecule the cross validation results is much better than that at each small group with a given protein site.
%2) The optimal effective distance in this test is 10 \AA,\ if all the data are ranked in the same query, which is consistent with that in the RF-Score. The detailed information of this 1322 complexes are provided in the supporting material.

\begin{figure}
\begin{center}
\begin{tabular}{cc}
\includegraphics[width=0.45\textwidth]{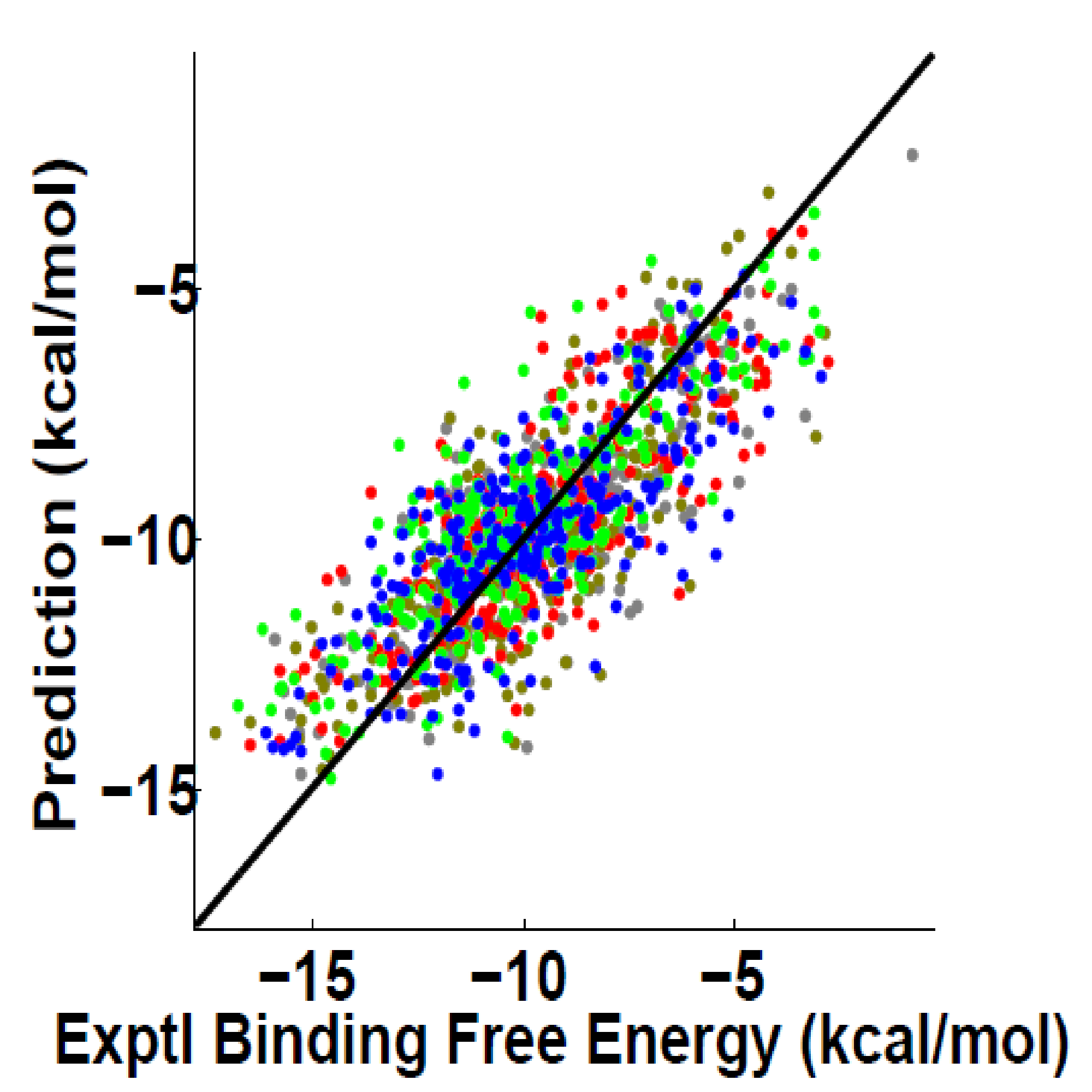}%{FiveFold_1322.png}
\includegraphics[width=0.55\textwidth]{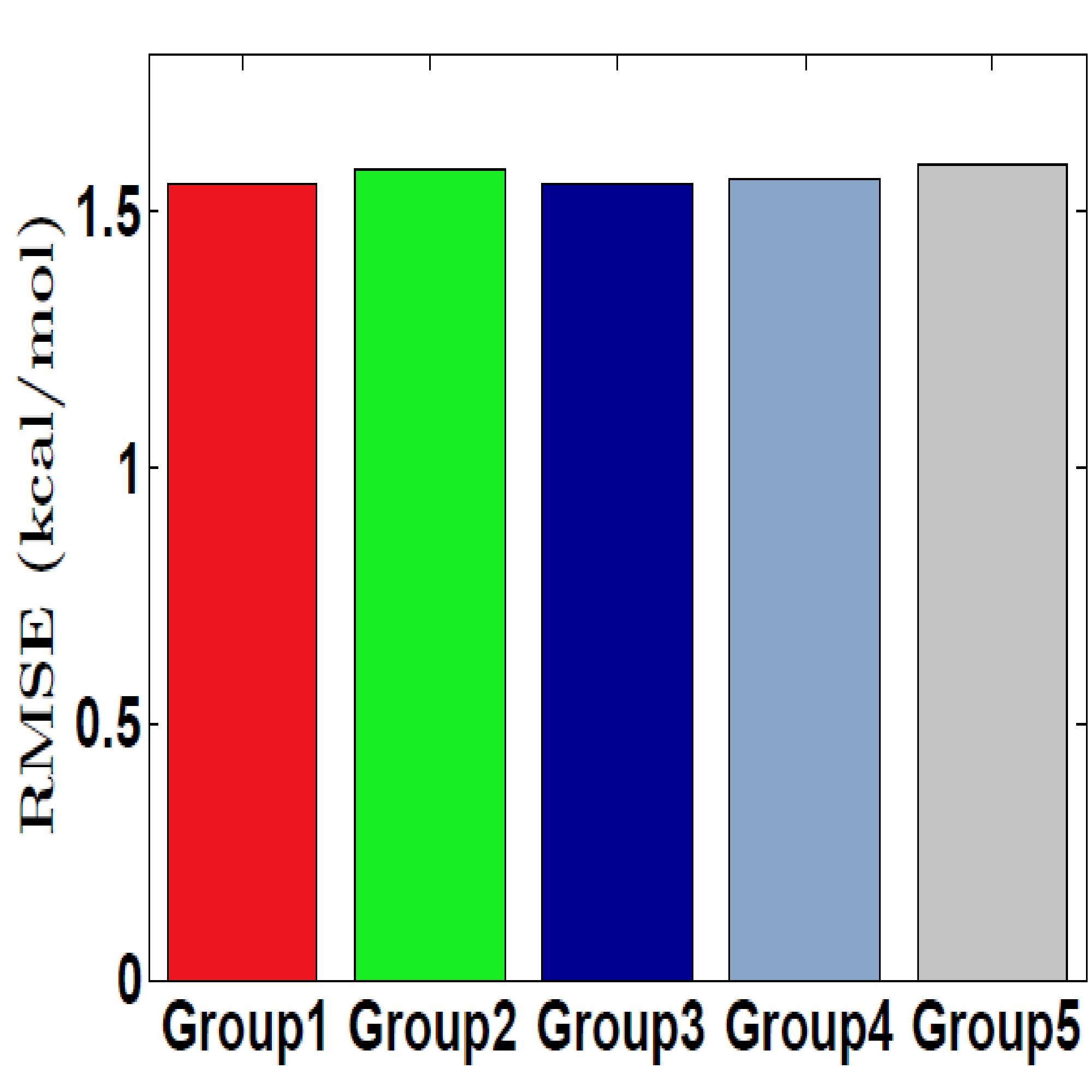}%{FiveFold1322_Err_EachGroup.png}
\end{tabular}
\end{center}
\caption{Five-fold cross validation on the     validation set  ($N=1322$). Left chart: correlation between experimental binding affinities and FFT-BP predictions. Right chart: RMSEs for five groups. Here,   RMSEs are 1.55,  1.58,  1.55,  1.56, and 1.59 kcal/mol for five groups, respectively. Overall Pearson correlation to the experimental binding affinities is 0.80.
}
\label{FiveFold1322}
\end{figure}

Finally, all the above predictions are based on the LTR ranking results. Alternatively, we can also carry out the prediction by using nearest neighbors and their associated features. We are interested to see the difference between these two approaches. To this end, we compute  the binding affinities of five-fold results  with different numbers of nearest neighbors and top features. Here top features are ranked by the LTR algorithm automatically according to their importance during the complex ranking. We list the top 50 important features to the protein ligand binding for the validation set in the Supporting material. We noted that the most important five features are the volume change, atomic Coulombic interaction of S atoms, area change of the C atoms in the protein and complex parts, and  electrostatic binding free energy.

\begin{table}
\centering
\caption{The RMSEs  (kcal/mol)  for  the   validation set  ($N=1322$) with different numbers of nearest neighbors and top features.  }
\begin{tabular}{c|cccccccccc}
\hline
%\multirow{2}{*}
{Number of }
 &\multicolumn{10}{c }{Number of top features}\\
% $\frac{{\rm Number of Features}}{{\rm Number of Nearest Neighbors}}$
  nearest neighbors &5 &10   &15   &20  &25   &30  &35  &40   &45   &50\\
\hline
1&   1.60&     1.60&   1.60&  1.61&  1.61&  1.61& 1.61& 1.61& 1.61& 1.62\\
2&   1.60&     1.60&   1.61&  1.61&  1.61&  1.61& 1.61& 1.62& 1.62& 1.62\\
3&   1.60&     1.59&   1.60&  1.70&  1.66&  1.68& 1.71& 1.70& 1.69& 1.70\\
4&   1.61&     1.57&   1.62&  1.71&  1.70&  1.72& 1.73& 1.70& 1.85& 1.83\\
5&   1.61&     1.60&   1.67&  1.74&  1.75&  1.74& 1.75& 1.73& 1.78& 1.77\\
6&   1.62&     1.61&   1.68&  1.79&  1.80&  1.81& 1.81& 1.88& 1.85& 1.85\\
7&   1.61&     1.61&   1.65&  1.78&  1.77&  1.78& 1.78& 1.81& 1.82& 1.82\\
8&   1.62&     1.62&   1.65&  1.74&  1.76&  1.76& 1.77& 1.78& 1.80& 1.81\\
9&   1.62&     1.61&   1.65&  1.74&  1.75&  1.76& 1.76& 1.76& 1.78& 1.77\\
10&  1.62&     1.62&   1.73&  1.74&  1.79&  1.80& 1.82& 1.82& 1.88& 1.90\\
\hline
\end{tabular}
\label{RMSerrorlist1322Test-adaptive}
\end{table}

The RMSEs of the tests with different numbers of top features and nearest neighbors involved are presented in Table \ref{RMSerrorlist1322Test-adaptive}. The optimal result is obtained when four nearest neighbor and 10 top features are utilized, with RMSE 1.57 kcal/mol. It is seen that when less than or equal to 10 top features are employed the prediction is quite accurate. However, with more features and more neighbors involved, the prediction become slightly worse. One possible reason is that the quality of the nearest neighbors is reduced when more neighbors are involved in the prediction. Indeed, the neighbors  that are not very close to the target molecule complex may make a  large difference to the prediction accuracy of the target complex. This issue also motivates us to seek a better set of features for protein-ligand binding analysis.

Figure \ref{FiveFold1322} depicts the optimal prediction results (Left chart) and RMSEs for each group (Right chart). It is seen that the RMSEs for all groups are almost  the same, indicating the unbiased nature of five-fold cross-validation. The success of proposed FFT-BP is implied by the small RMSEs of 1.55 $\sim$ 1.59 kcal/mol and the high overall Pearson correlation of 0.80.

\subsubsection{Validation on the training set ($N=3589$)}

We also consider   the five-fold cross validation  on our training set of 3589 complexes. We randomly divide this data set into five groups with 717, 718, 718, 718, and 718 complexes, respectively. In the five-fold cross validation, each time we regard one group of molecules as the test set without binding affinity data, and using the remaining four groups to predict the binding affinities of the selected test set.

\begin{table}
\centering
\caption{The RMSEs (kcal/mol) for the five-fold cross validation on the  training set  ($N=3589$) with different number of nearest neighbors and top features.  }
\begin{tabular}{c|cccccccccc}
\hline
%$\frac{{\rm Number of Features}}{{\rm Number of Nearest Neighbors}}$
%\multirow{2}{*}
{Number of } &\multicolumn{10}{c }{Number of top features}\\
  nearest neighbors   &5 &10   &15   &20  &25   &30  &35  &40   &45   &50\\
\hline
1&   2.00&     1.99&   2.00&  2.00&  2.00&  2.01& 2.01& 2.01& 2.01& 2.02\\
2&   2.00&     1.99&   2.00&  1.99&  2.00&  2.01& 2.01& 2.01& 2.01& 2.01\\
3&   2.01&     2.00&   2.00&  2.00&  2.00&  2.00& 2.02& 2.01& 2.01& 2.01\\
4&   2.00&     2.01&   2.00&  1.99&  2.00&  2.01& 2.01& 2.01& 2.02& 2.01\\
5&   2.01&     2.00&   2.01&  2.01&  2.01&  2.01& 2.01& 2.01& 2.01& 2.02\\
6&   2.00&     1.99&   1.99&  2.00&  2.00&  2.01& 2.01& 2.01& 2.01& 2.01\\
7&   2.00&     2.00&   2.00&  2.00&  2.01&  2.01& 2.02& 2.02& 2.02& 2.02\\
8&   2.00&     1.99&   1.98&  1.99&  1.99&  2.00& 2.00& 2.00& 2.01& 2.00\\
9&   2.00&     2.00&   2.00&  2.01&  2.02&  2.05& 2.05& 2.05& 2.05& 2.04\\
10&  1.99&     2.00&   2.00&  2.03&  2.04&  2.07& 2.08& 2.08& 2.08& 2.08\\
\hline
\end{tabular}
\label{PDB2015Test-adaptive}
\end{table}

\begin{figure}
\begin{center}
\begin{tabular}{cc}
\includegraphics[width=0.45\textwidth]{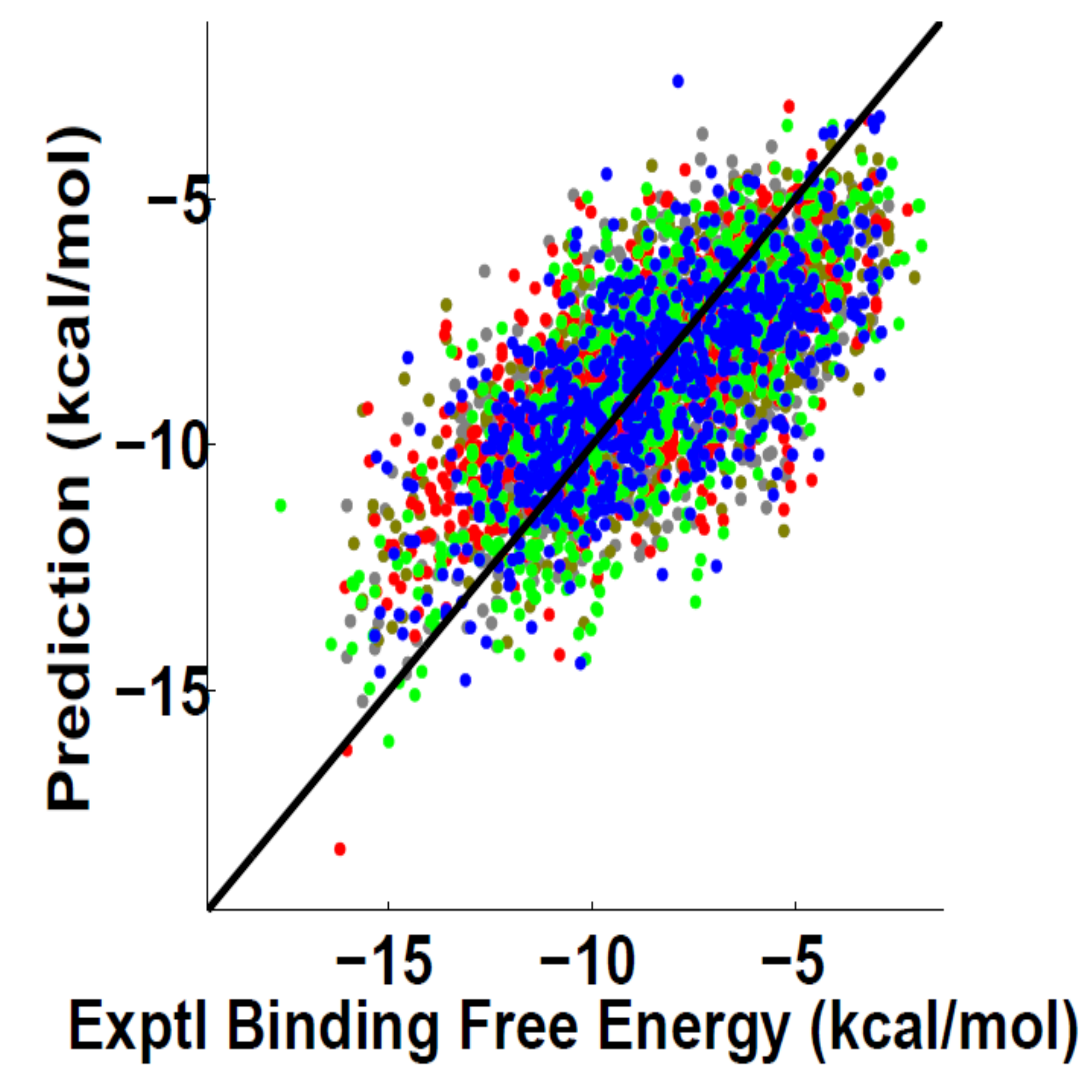}%{Refined2015_FiveFold.png}
\includegraphics[width=0.55\textwidth]{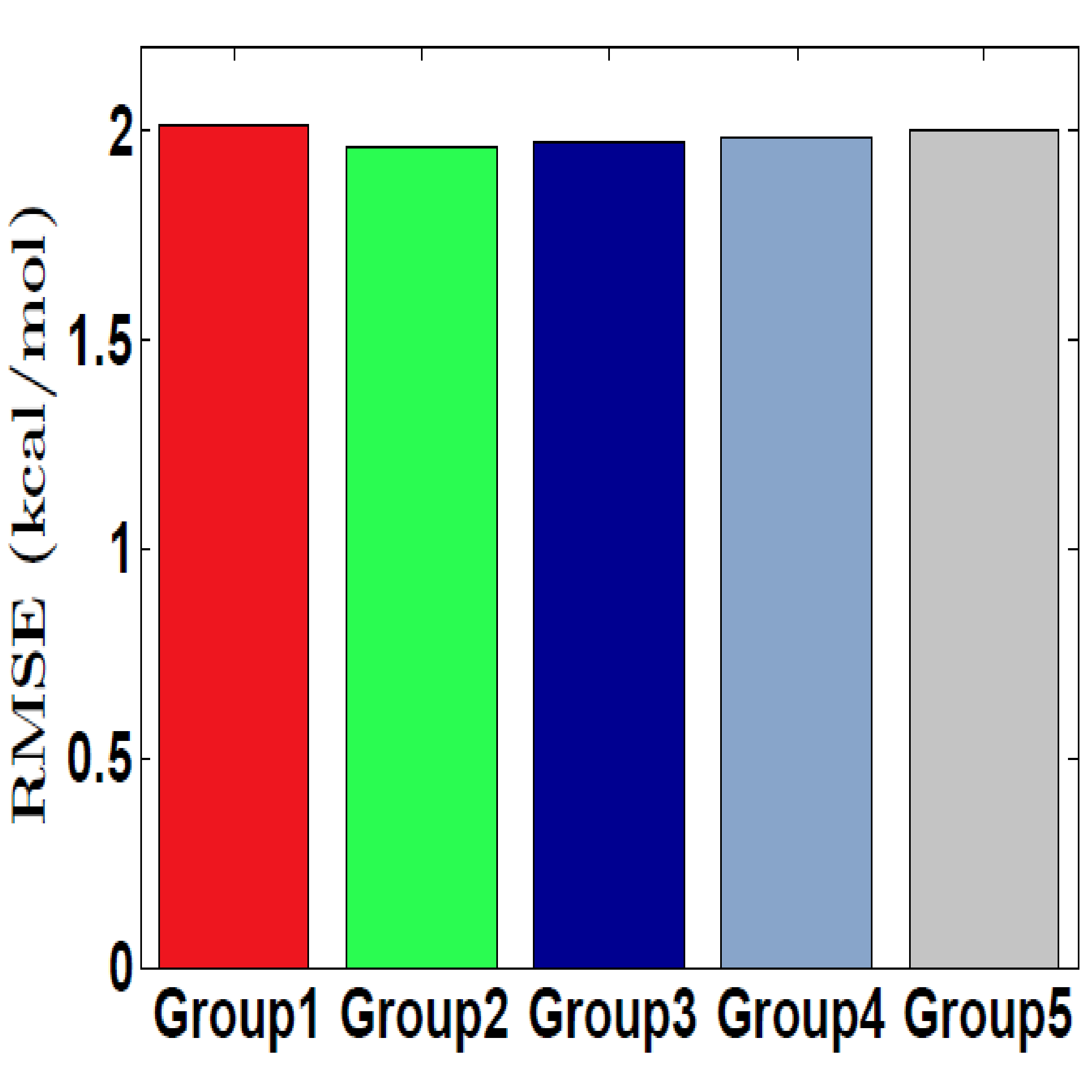}%{Refined2015_Err_EachGroup.png}
\end{tabular}
\end{center}
\caption{Five-fold cross validation on the training set (3589 complexes). Left chart: correlation between experimental binding affinities and FFT-BP predictions. Right chart: RMSEs for five groups. Here,  RMSEs are 2.01,  1.96,  1.97,  1.98, and 2.00 kcal/mol for five groups, respectively. Overall Pearson correlation to the experimental is 0.70.
}
\label{FiveFold2015}
\end{figure}

 Directly using the ranking score as the predicted binding affinity leads to RMSE 2.00 kcal/mol. Alternatively, we can predict binding affinities using the nearest neighbors and top features.

Table \ref{PDB2015Test-adaptive} shows the RMSEs for the five-fold cross validation test on the training set ($N=3589$). The number of nearest neighbors is varied from 1 to 10 and the number of to features is changed from 5 to 50. The most important 50 features indicated from the LTR algorithm  are provided in the Supporting material. Five top  important features  are volume change, electrostatics binding free energy, and van der Waals interactions between C-S, C-O and C-N pairs, respectively. The optimal prediction is achieved when 8 nearest neighbors and top { 15} features are used for binding affinity prediction, with the RMSE being 1.98 kcal/mol. Different numbers of nearest neighbors and top features basically give very consistent predictions. Compared to the five-fold test on the 1322 protein ligand complexes, the prediction errors on this set are much larger, which is partially due to the fact that structures in this test set is more complexes. For example, binding-site metal effects are presented without an appropriate treatment. We believe a better treatment of metal effects and a classification of  ligand molecules would improve the FFT-BP prediction.

Figure \ref{FiveFold2015} depicts the optimal prediction results (Left chart) and RMSEs for each group (Right chart). These tests demonstrate the following two facts. First, five-fold cross validation prediction is unbiased. The prediction results do not depends on the data itself and the RMSEs for all groups are almost at the same level. Second, when the protein-ligand complexes  become diverse, the prediction becomes slightly worse due to the lack of similar complexes for certain clusters.
%than limit protein binding site case. On the other hand, this reflects the basic principle of the machine learning algorithms, with more available and reliable data, we can build better model from data itself.

\subsection{Blind predictions on three test sets}

To further verify the  accuracy  of the FFT-BP, we perform the blind prediction on three benchmark test sets. The training set ($N=3589$) that is processed from the PDBBind v2015 refined set is utilized for the training in blind predictions { of the benchmark set of 100 complexes and PDBBind v2015 core set}. { In addition, the training set $(N=1082)$ processed from PDBBind v2007 refined set is employed as the training data in a blind prediction of PDBBind v2007 core set}. Due to the LTR algorithm used in our FFT-BP, the RMSE and correlation of our FFT-BP prediction would be around 0 kcal/mol and 1, respectively, had we include all the test set complexes in our training set. Therefore, in each blind prediction,   we carefully exclude the overlapping test set complexes from the training set and re-train the training set  with a reduced number of complexes.
% the complexes overlapped with the benchmark test sets are ruled out is used to train the FFT-Score. Subsequently, the obtained scoring function is used to predict the binding affinity on the benchmark sets.

\subsubsection{Prediction on the benchmark set ($N=100$)}

\begin{figure}[!ht]
\small
\centering
\includegraphics[width=6.5cm,height=5.5cm]{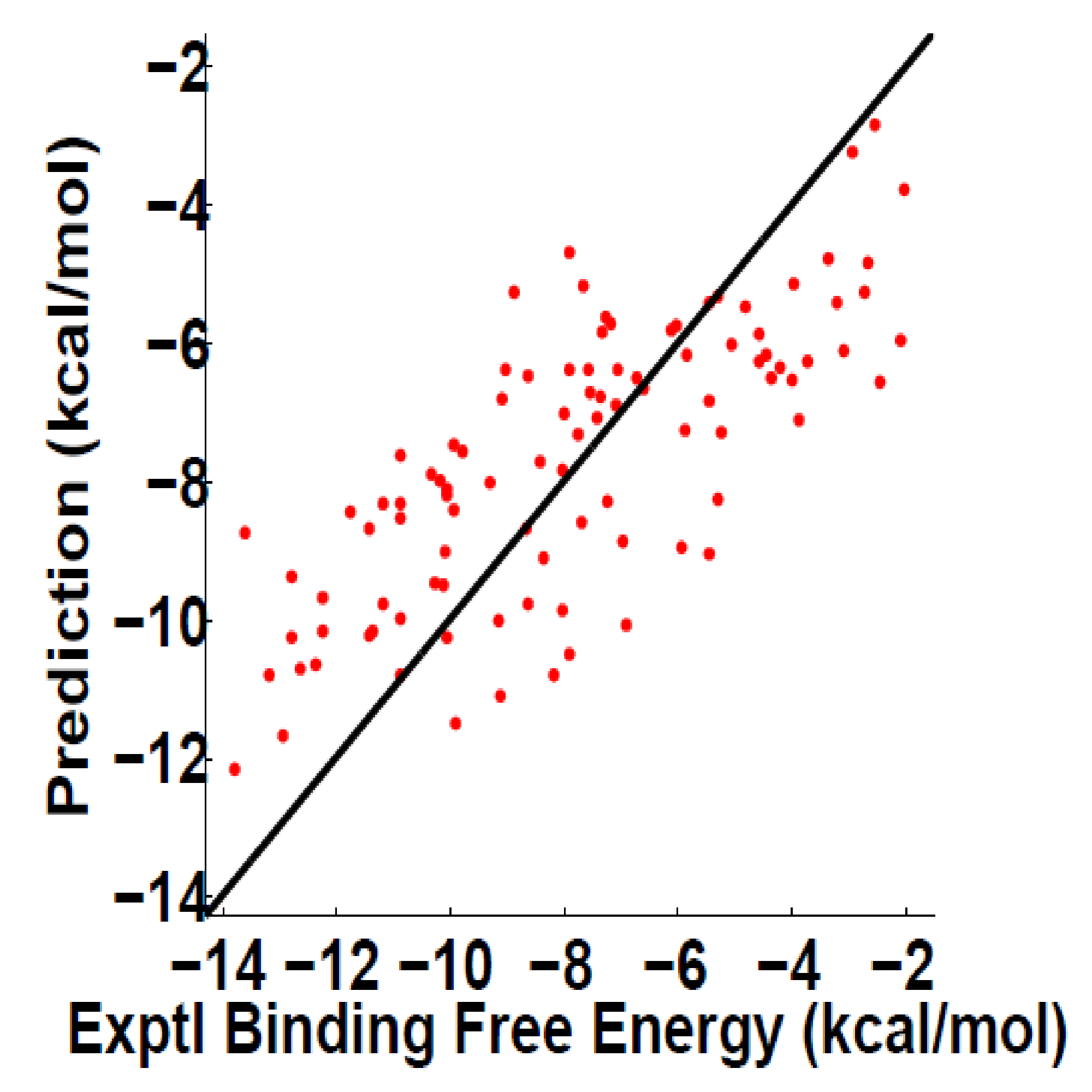}%{Benchmark_100.png}
\caption{The correlation   between experimental binding free energies and FFT-BP predictions  on the benchmark test set ($N= 100$) with the RMSE of  1.99 kcal/mol and the Pearson correlation  of 0.75.}
\label{Benchmark100-Err}
\end{figure}

\begin{table}
\centering
\caption{The   RMSEs  (kcal/mol)  of the FFT-BP for  the benchmark test set ($N= 100$) with different numbers of nearest neighbors and top features.}
\begin{tabular}{c|cccccccccc}
\hline
%$\frac{{\rm Number~ of ~Features}}{{\rm Number ~of~ Nearest Neighbors}}$
%\multirow{2}{*}
{Number of  } &\multicolumn{10}{c }{Number of top features}\\
 nearest neighbors &5 &10   &15   &20  &25   &30  &35  &40   &45   &50\\
\hline
1&   2.00&     2.00&   2.00&  2.00&  2.00&  2.00& 2.00& 2.00& 2.00& 2.00\\
2&   2.01&     2.01&   1.99&  1.99&  2.01&  2.00& 2.01& 2.01& 2.01& 2.01\\
3&   2.00&     2.00&   2.00&  2.00&  2.00&  2.00& 2.00& 2.01& 2.01& 2.01\\
4&   2.01&     2.01&   2.01&  2.00&  2.00&  2.00& 2.01& 2.01& 2.01& 2.01\\
5&   2.01&     2.01&   2.01&  2.01&  2.01&  2.01& 2.00& 2.01& 2.01& 2.01\\
6&   2.02&     2.01&   2.01&  2.01&  2.01&  2.01& 2.01& 2.01& 2.01& 2.01\\
7&   2.02&     2.02&   2.01&  2.01&  2.01&  2.01& 2.01& 2.01& 2.01& 2.01\\
8&   2.01&     2.01&   2.01&  2.01&  2.01&  2.01& 2.02& 2.02& 2.02& 2.02\\
9&   2.01&     2.01&   2.01&  2.01&  2.01&  2.01& 2.02& 2.01& 2.01& 2.01\\
10&  2.01&     2.01&   2.01&  2.02&  2.01&  2.02& 2.01& 2.02& 2.02& 2.02\\
\hline
\end{tabular}
\label{Test100-adaptive}
\end{table}

First of all, we consider a popular benchmark set  originally used by Wang {\it et al} \cite{RenxiaoWang:2003CompareSF}. This set contains 100 protein ligand complexes which involves a large variety of protein receptors. Originally this test set was used to test the performance of a large amount of well-known scoring functions and docking algorithms \cite{RenxiaoWang:2003CompareSF}. Recently, Zheng {\it et al } have utilized this test set to demonstrate the superb performance of their KECSA method \cite{Zheng:2013KECSA}. In this work, we examine the accuracy and robustness of  our FFT-BP on this benchmark test set.

\begin{figure}[!ht]
\small
\centering
\includegraphics[width=12cm,height=12.5cm]{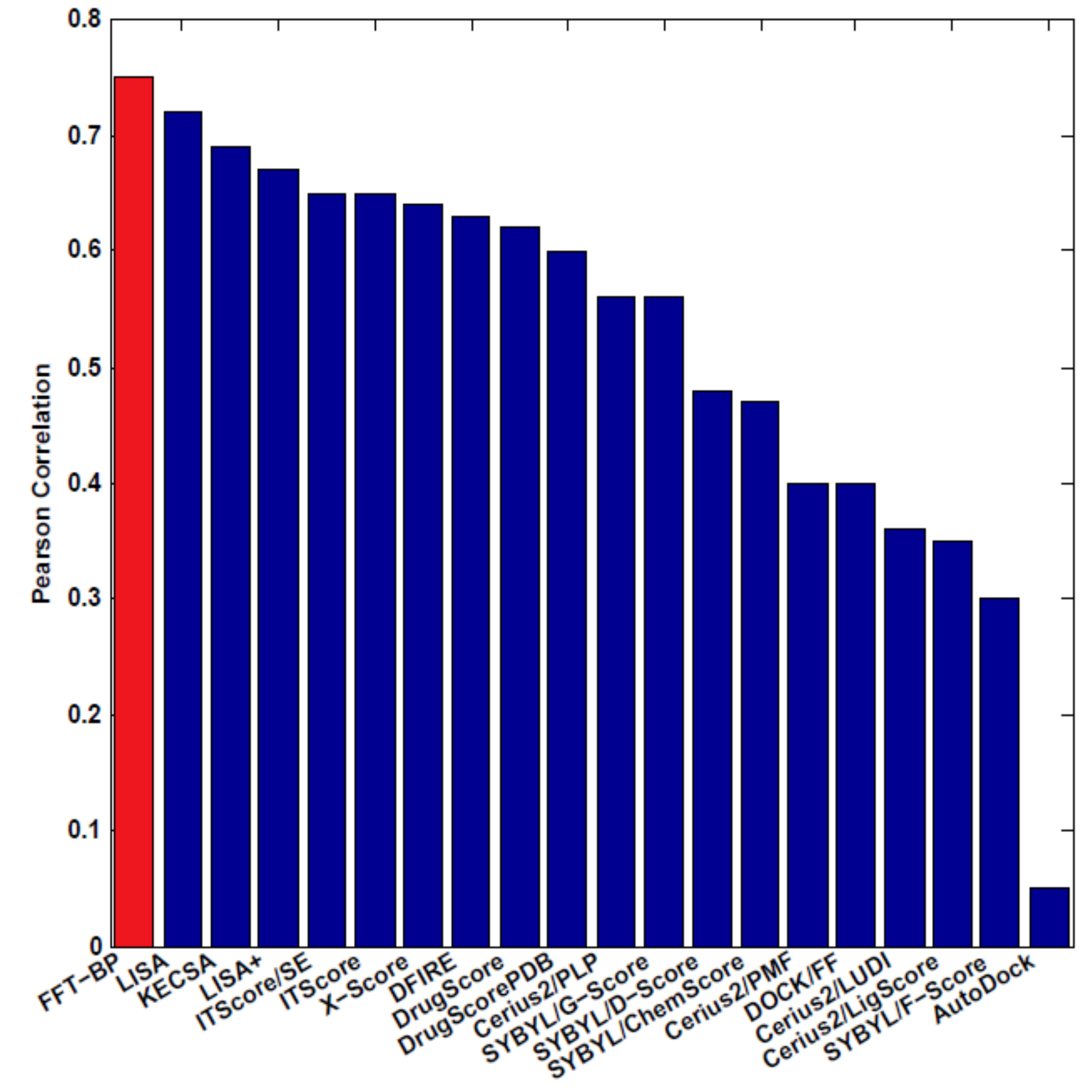}%{Compare_100.png}
\caption{Performance comparison between different scoring functions on  the benchmark test set ($N= 100$). The binding affinity comparisons was done for FFT-BP, and 19 well-known scoring functions, namely LISA, KECSA, LISA+ \cite{Zheng:2013KECSA, Zheng:2015LISA}, ITScore/SE \cite{ITScoreSE:2010}, ITScore \cite{ITScore:2006}, X-Score \cite{WangRenXiao:2002}, DFIRE \cite{DFIRE:2005}, DrugScoreCSD \cite{DrugScoreCSD:2005}, DrugScorePDB \cite{DrugScorePDB:2000}, Cerius2/PLP \cite{Cerius2:PLP1}, SYBYL/G-Score \cite{G-Score}, SYBYL/D-Score \cite{D-Score}, SYBYL/ChemScore \cite{Eldridge:1997}, Cerius2/PMF \cite{PMFScore:1999}, DOCK/FF \cite{D-Score}, Cerius2/LUDI \cite{CERIUS2LUDI}, Cerius2/LigScore \cite{CERIUS2:2000}, SYBYL/F-Score \cite{F-Score}, and AutoDock \cite{AutoDock:1998}.}
\label{Compare100}
\end{figure}

%Consistent different number involved, also consistent with five fold cross validation

Directly using the ranking score as the predicted binding affinity leads to the RMSE of 2.01 kcal/mol and Pearson correlation coefficient of 0.75. Alternatively, we examine FFT-BP predictions using different numbers of nearest neighbors and top features. Table \ref{Test100-adaptive} lists the predicted RMSEs for the  benchmark  set ($N=100$). The numbers of nearest neighbors and tops features vary from 1 to 10 and from 5 to 50, respectively.  The most important 50 features indicated by the LTR algorithm  are provided in the Supporting material.  Five top important features are volume change, electrostatics binding free energy, van der Waals interaction between C-S and C-C pairs, and the complex's area change. The optimal prediction is reached when 2 nearest neighbors and top 15 or 20 features are used for binding prediction. The corresponding   RMSEs and correlation coefficients for both cases are 1.99 kcal/mol and 0.75, respectively. Different numbers of nearest neighbors and top features basically give rise to very consistent predictions. We also note that the prediction errors for this 100 test set are very similar to those of the five-fold cross validation tests on our training set ($N=3589$). This consistency indicates the robustness of the proposed FFT-BP in binding affinity predictions.

Figure \ref{Benchmark100-Err} illustrates the optimal prediction results compared to the experimental data. The RMSE and Pearson correlation coefficient are 1.99 kcal/mol and 0.75, respectively. This test set is a critical test set with diverse protein-ligand complexes and a wide range of experimental binding free energies. In our prediction, most predictions are quite appealing with less than 2 kcal/mol RMSE compared to  experimental results.

Many outstanding scoring functions have been tested on this test set as summarized by Zheng {\it et al} \cite{Zheng:2013KECSA}. Here we also add our prediction to this list.  As shown in Fig. \ref{Compare100}, the performance of our FFT-BP   is highlighted with  red color. The performance of  other 19 scoring functions are due to the courtesy of Ref. \cite{Zheng:2013KECSA}.

\subsubsection{Prediction on the PDBBind v2007 core set ($N=195$)}

\begin{figure}[!ht]
\small
\centering
\includegraphics[width=6.5cm,height=5.5cm]{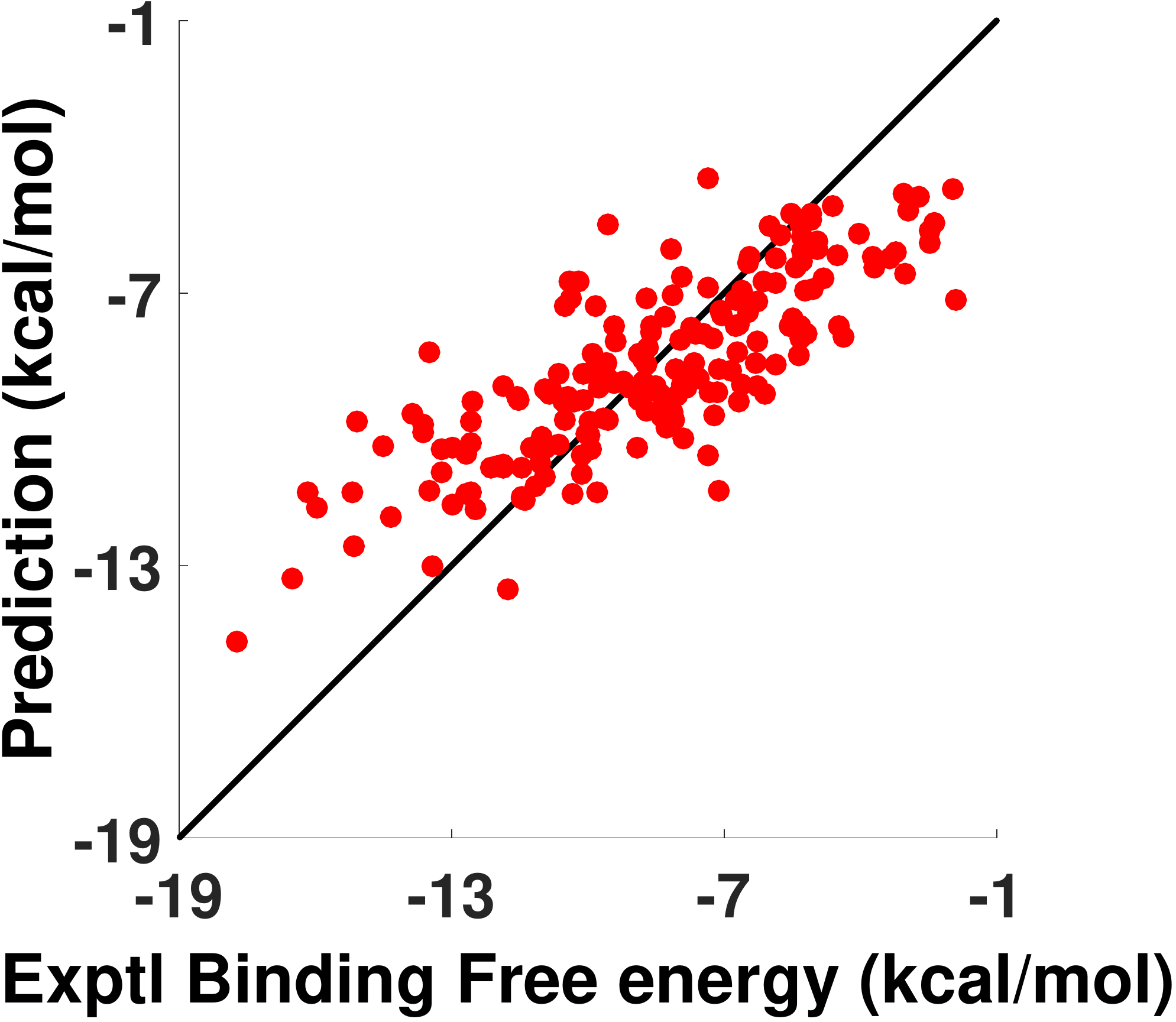}%{Core2007.png}
\caption{ {The correlation of between experimental binding free energies and FFT-BP predictions  on the PDBBind core 2007 ($N=195$). The RMSE and Pearson correlation coefficient are 2.02 kcal/mol and 0.80, respectively.}}
\label{Pred2007}
\end{figure}

%Figure \ref{Pred2007} illustrates the correlation between experimental binding free energies and the  best predictions obtained by the FFT-BP. Obviously, there is a bias in the predicted  binding affinities, which will be addressed in our future work.

\begin{figure}[!ht]
\small
\centering
\includegraphics[width=12cm,height=12.5cm]{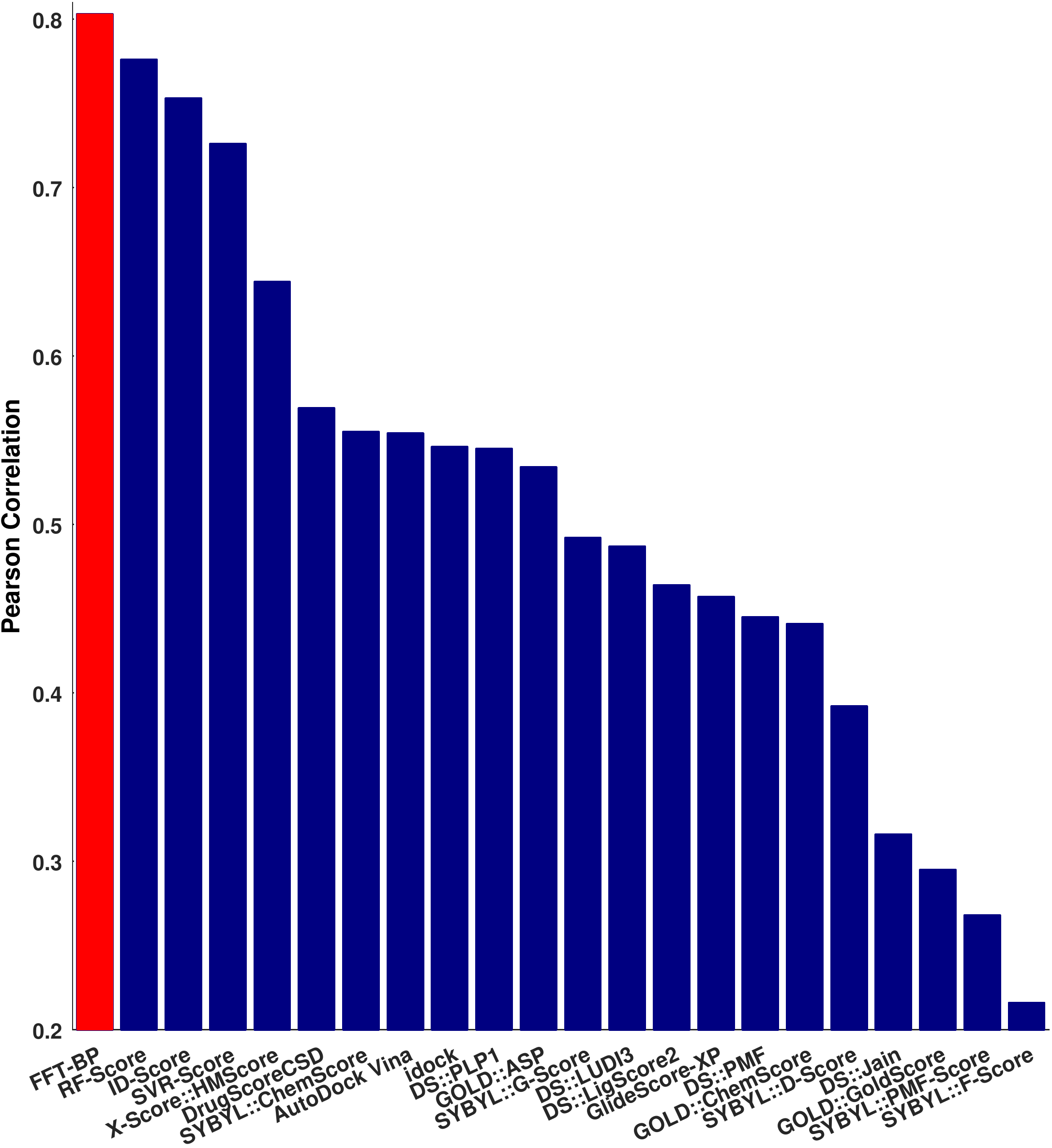}%{Core2007_Compare.png}
\caption{ {Performance comparison between different scoring functions on the PDBBind v2007 core set ($N=195$).  The performances of the other scoring function are adopted  from the literature \cite{istar:2014, Pedro:2010Binding, IDScore:2013, Pedro:2012New, RenxiaoWang:2009Compare}.}}.
\label{Core2007-Compare}
\end{figure}

PDBBind v2007 core set ($N=195$) which contains high quality data  mainly aims for testing the performance of scoring functions \cite{PDBBind:2015}. It  has been employed to study and compare  many excellent scoring functions \cite{istar:2014, Pedro:2010Binding, IDScore:2013, Pedro:2012New, RenxiaoWang:2009Compare}. { To predict the binding affinity of this core set, it is a rational to employ the PDBBind v2007 refined set instead of v2015 one as the training set. Definitely, the training set here will not overlap with the test set. The score from the MART machine learning method is directly used for the prediction. Figure \ref{Pred2007} illustrates the correlation between experimental binding free energies and the  best predictions obtained by the FFT-BP. The Pearson correlation coefficient and RMSE by FFT-BP are, respectively, 0.80 and 2.03 kcal/mol. }

Li {\it et al} have given a comparison of tests on the PDBBind v2007 core set ($N=195$) using many outstanding scoring functions \cite{istar:2014}. In this regarding,  we also plot the performance  of our FFT-BP   in terms of  Pearson correlation coefficient  in Fig. \ref{Core2007-Compare}. The FFT-BP correlation coefficient of { 0.80}  is highlighted with  red color.

\subsubsection{Prediction on the PDBBind v2015 core set ($N=195$)}

\begin{figure}[!ht]
\small
\centering
\includegraphics[width=6.5cm,height=5.5cm]{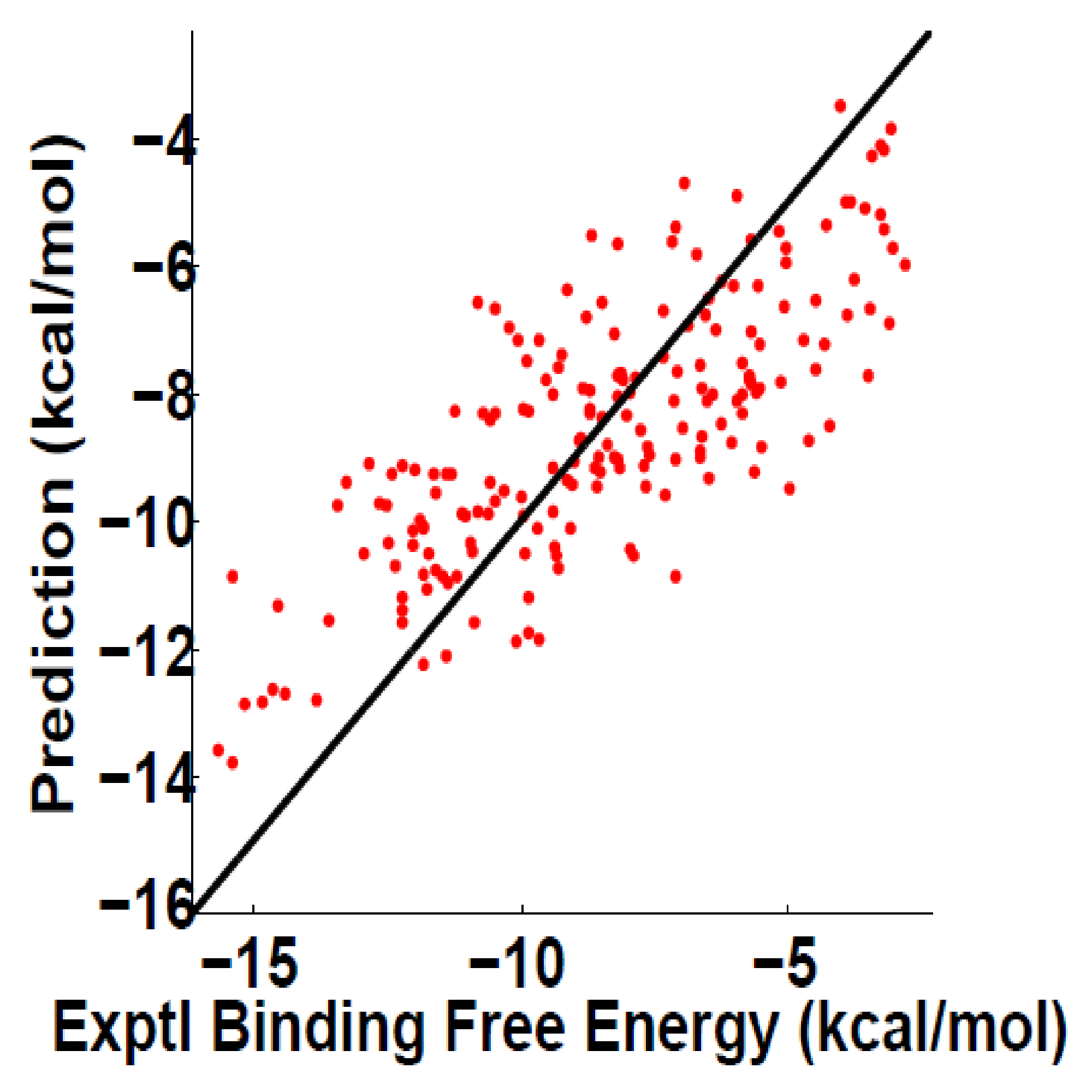}%{Core2015.png}
\caption{The correlation  between experimental binding free energies and FFT-BP predictions  on the PDBBind v2015 core set  ($N=195$). The RMSE and Pearson correlation coefficient are 1.92 kcal/mol and 0.78, respectively.}
\label{Pred2015}
\end{figure}

\begin{table}
\centering
\caption{The prediction RMSEs  (kcal/mol)  for the PDBBind v2015 core set ($N=195$) with different numbers of nearest neighbors and top features.  }
\begin{tabular}{c|cccccccccc}
\hline
% $\frac{{\rm Number of Features}}{{\rm Number of Nearest Neighbors}}$
%\multirow{2}{*}
{Number of  } &\multicolumn{10}{c }{Number of top features}\\
 nearest neighbors &5 &10   &15   &20  &25   &30  &35  &40   &45   &50\\
\hline
1&   1.95&     1.95&   1.95&  1.95&  1.95&  1.95& 1.95& 1.95& 1.95& 1.95\\
2&   1.95&     1.94&   1.95&  1.95&  1.95&  1.95& 1.95& 1.95& 1.96& 1.96\\
3&   1.94&     1.94&   1.95&  1.95&  1.95&  1.95& 1.95& 1.95& 1.95& 1.95\\
4&   1.94&     1.94&   1.93&  1.93&  1.94&  1.94& 1.94& 1.95& 1.95& 1.95\\
5&   1.95&     1.95&   1.92&  1.94&  1.95&  1.95& 1.96& 1.95& 1.95& 1.95\\
6&   1.95&     1.95&   1.95&  1.95&  1.96&  1.96& 1.95& 1.96& 1.95& 1.94\\
7&   1.95&     1.93&   1.93&  1.94&  1.95&  1.97& 1.95& 1.95& 1.95& 1.95\\
8&   1.95&     1.95&   1.95&  1.96&  1.96&  1.97& 1.97& 1.96& 1.95& 1.95\\
9&   1.94&     1.94&   1.94&  1.94&  1.94&  1.95& 1.95& 1.94& 1.94& 1.94\\
10&  1.95&     1.95&   1.95&  1.95&  1.95&  1.94& 1.94& 1.94& 1.94& 1.94\\
\hline
\end{tabular}
\label{Test2015-adaptive}
\end{table}

\begin{figure}[!ht]
\small
\centering
\includegraphics[width=12cm,height=12.5cm]{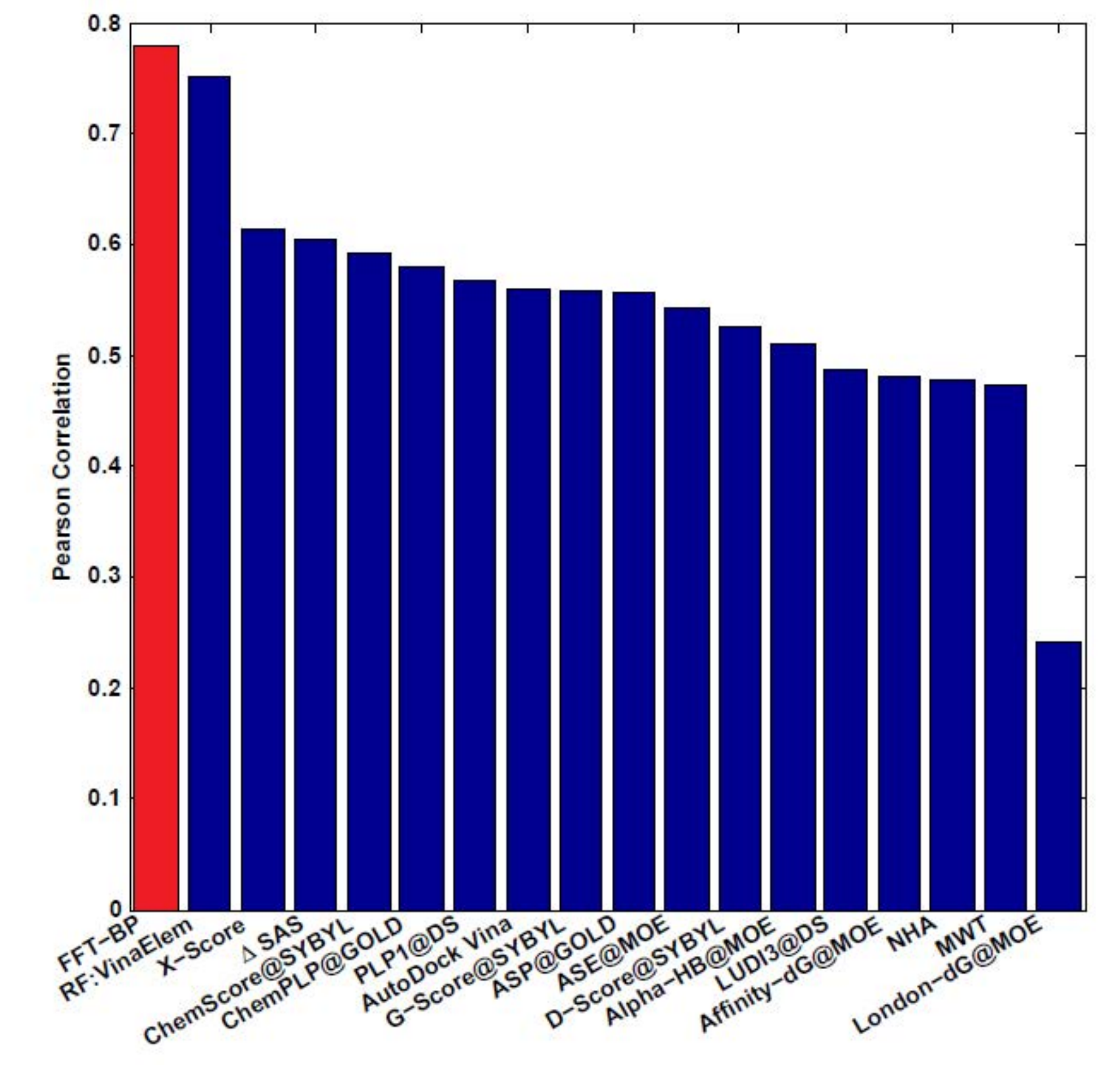}%{Core2015_compare.png}%{Core2007_Compare.png}
\caption{{ Performance comparison between different scoring functions on the PDBBind v2013 core set ($N=195$).  The performances of the other scoring function are adopted  from the literature \cite{YLi:2014,HongJianLi:2015}.}}
\label{Core2015-Compare}
\end{figure}

Finally, we perform a test on the  PDBBind v2015 core set ($N=195$), which contains  high quality  experimental data. { PDBBind v2015 core set is the same as  PDBBind v2013  core set and  PDBBind v2014 core set.} 
This test set is also quite challenging  due its  diversity of 65 protein-ligand clusters and  a wide binding affinity range. In a similar routine, we first consider the FFT-BP prediction with different numbers of neighbors and top features. Table \ref{Test2015-adaptive} shows the  RMSEs of FFT-BP for PDBBind v2015 core  set ($N=195$). The top   50 features are also listed in the Supporting material. The most important features are similar to those in previous tests, which indicates that the volume change, electrostatic binding free energy and van der Waals interactions are of fundamental importance to the protein-ligand binding. It is worth noting that the RMSEs of FFT-BP predictions are lower than those from  earlier test sets.   A possible reason is that this data set is consistent with the training set as  both obtained from the PDBBind 2015 refined set. Additionally,  a  better data quality might also contribute our better predictions.   Our optimal prediction has the RMSE of 1.92 kcal/mol and Pearson correlation coefficient of 0.78, when 5 nearest neighbors and 15 features are used for the prediction.

%%Table \ref{Test2015-adaptive} shows the prediction RMSEs for PDBBind2007 core test set, the number of nearest neighbors and tops features involved ranging from 1 to 10 and 5 to 50, respectively. The most important 50 features that LTR algorithm indicates are provided in the supporting material, the top five important features among them are: volume change; electrostatics binding free energy; van der Waals interaction between C-O; the complex's area change; and van der Waals interaction between C-C. These features basically consistent with all the previous test cases. The optimal prediction achieves when 6 nearest neighbors and top 15 or 30 features are used for binding prediction, the RMSEs for both cases are 2.08 kcal/mol For this test, the prediction based on different number of nearest neighbors and features do not differ much from each other.

Figure \ref{Pred2015} plots the  correlation  between experimental binding free energies and FFT-BP predictions  on the   PDBBind v2015 core set  ($N=195$). Compared to the earlier two blind predictions, the prediction on this set is more accurate. However, similar to the behavior in two other test sets, the present prediction is biased. This issue will be studied in our future work.  { Note that  PDBBind v2015 core set is the same as the  PDBBind v2013 core set, which has many test results  \cite{YLi:2014,HongJianLi:2015}. For a comparison, we plot the performance of our scoring function against several existing famous scoring functions  \cite{YLi:2014,HongJianLi:2015}, as illustrated in Fig. \ref{Core2015-Compare}.}  

\section{Concluding remarks}\label{Conclusion}

In this work, we  propose   a new scoring function, feature functional theory - binding predictor (FFT-BP). FFT-BP is constructed based on three fundamental assumptions, namely, representability, feature-function relationship, and similarity assumptions.  A validation set of 1322 complexes, { two training sets with 3589 complexes (PDBBind v2015 refined set) and 1085 complexes (PDBBind v2007 refined set)}, and three test sets with 100, 195 and 195 complexes are considered in the present work  to validate the proposed method, explore its utility, demonstrate its performance and reveal its deficiency.  Extensive numerical experiments indicate that FFT-BP delivers some of the most accurate blind predictions   in the field with the root-mean-square error being around 2 kcal/mol and Pearson   correlation coefficient being around 0.76.

A major advantage of FFT-BP is that it extracts   microscopic features from conventional  implicit solvent models so that the validity of these physical models for binding analysis and prediction can be systematically examined. Consequently, the proposed FFT-BP can be improved via the improvement of our understanding on physical models. Another advantage of FFT-BP is that it provides a framework to systematically incorporates and continuously absorb advanced machine learning algorithms to improve its predictive power. The other advantage of FFT-BP is that it becomes more and more accurate as the existing binding database becomes  larger and larger.

This work is our first attempt in exploring the mathematical modeling of the protein-ligand binding affinity. Our model can be further improved in several aspects. First,  we have employed a very crude force field parametrized of the Poisson model. More accurate Poisson-Boltzmann (PB) modeling, such as polarizable PB model,  and feature extraction from more accurate quantum mechanics/molecular mechanics (QM/MM) models will improve the present FFT-BP. Additionally,  we employ the MART algorithm for the molecules ranking. More sophisticated machine learning algorithms, such as  deep learning, can potentially improve    FFT-BP prediction, and eliminate the current prediction bias in test sets. Finally, a deficiency of the current model is that it neglects  the metal effect on protein-ligand binding affinity.   The incorporation of this effect into our model is under our investigation.

\section*{Acknowledgments}
This work was supported in part by NSF Grant   IIS- 1302285 and MSU Center for Mathematical Molecular Biosciences Initiative. We thank  Emil Alexov, Michael Gilson, Ray Luo, Wei Yang and John Zheng  for useful discussions.

%
%\bibliographystyle{abbrv}
%%\bibliographystyle{plain}
%\bibliography{refs}
\end{document}